\newif\ifspringer
\springerfalse		

\ifspringer  
\documentclass[runningheads]{springer_format/llncs}

\usepackage[T1]{fontenc}

\usepackage{newtxtext}       \usepackage[varvw]{newtxmath}

\usepackage{graphicx}
 \else
\documentclass[11pt]{article}
\fi

\ifspringer \else \usepackage[letterpaper,margin=1in] {geometry} \fi
\usepackage[utf8]{inputenc}

\usepackage{enumerate}
\usepackage{enumitem}
\setlist[enumerate,1]{label=(\roman*), leftmargin=2.2em, itemsep=0pt,topsep=0.3em}
\setlist[enumerate,2]{label=(\alph*),nosep}
\setlist[itemize]{itemsep=0pt,topsep=0.3em}

\usepackage[table,dvipsnames]{xcolor}
\definecolor{linkcol}{rgb}{0,0,0.38}
\definecolor{citecol}{rgb}{0,0.2,0}
\definecolor{urlcol}{rgb}{0.1,0.35,0}
\usepackage[pdfpagelabels,bookmarks=false,hyperfootnotes=false]{hyperref}
\hypersetup{colorlinks, linkcolor=linkcol, citecolor=citecol, urlcolor=urlcol}
\usepackage{url}

\ifspringer\else \usepackage{amsthm} \fi \usepackage{amsmath}
\ifspringer\else \usepackage{amssymb} \fi 

\ifspringer \spnewtheorem{openquestion}{Open question}{\bfseries}{\rmfamily} \else\fi 

\usepackage{thmtools}
\usepackage{thm-restate}
\usepackage{mathtools}
\usepackage{bbm}
\usepackage{xfrac}

\usepackage[nameinlink, capitalise]{cleveref}

\crefname{appendix}{Appendix}{Appendices}
\Crefname{appendix}{Appendix}{Appendices}

\ifspringer\else    \newtheorem{theorem}{Theorem}[section]
\newtheorem{lemma}[theorem]{Lemma}
\newtheorem{proposition}[theorem]{Proposition}

\theoremstyle{definition}
\newtheorem{definition}[theorem]{Definition}

\newtheorem{problem}[theorem]{Problem}
\newtheorem{openquestion}[theorem]{Open Question}

\fi

\ifspringer \AddToHook{env/proof/end}{\qed}
\else\fi

\usepackage[linesnumbered, ruled, vlined]{algorithm2e}
\usepackage{aligned-overset}
\usepackage[normalem]{ulem}
\usepackage{xspace}

\usepackage[
backend=biber,
style=alphabetic,
citestyle=alphabetic,
maxalphanames=5,
maxcitenames=99,
mincitenames=98,
maxbibnames=99,
giveninits=true,
]{biblatex}

\usepackage{tikz}
\usetikzlibrary{calc}
\usetikzlibrary{math}
\usetikzlibrary{shapes.geometric}
\usetikzlibrary{arrows}
\usetikzlibrary{decorations.pathreplacing, decorations.pathmorphing}
\usepackage{float}
\usepackage{graphicx}
\usepackage[margin=10pt,font=small,labelfont=bf,skip=-5pt]{caption}
\usepackage[subrefformat=parens]{subcaption}

\usepackage[textsize=footnotesize, color=orange!40]{todonotes}
\let\originaltodo\todo
\renewcommand{\todo}[1]{\PackageWarning{mytodos}{TODO found: #1}\originaltodo{#1}}

\usepackage[most]{tcolorbox}
\newtcolorbox{commentbox}{
  colback=blue!5,        colframe=blue!50!black, boxrule=0.8pt,
  arc=4pt,               left=6pt, right=6pt, top=6pt, bottom=6pt,
  fonttitle=\bfseries,
  title=Comment
}

\newcommand\bR{\ensuremath{\mathbb{R}}}

\newcommand\cI{\ensuremath{\mathcal{I}}}

\newcommand\cL{\ensuremath{\mathcal{L}}}

\def\cupp{\stackrel{.}{\cup}}
\def\bigcupp{\stackrel{.}{\bigcup}}

\DeclarePairedDelimiter\ceil{\lceil}{\rceil}
\DeclarePairedDelimiter\floor{\lfloor}{\rfloor}

\DeclareMathOperator*{\supp}{supp}

\newcommand\parset[1]{\ensuremath{\left\{ #1 \right\}}}

\DeclareRobustCommand{\OPT}{\ensuremath{\mathrm{OPT}}}
\DeclareRobustCommand{\LP}{\ensuremath{\mathrm{LP}}}

\DeclareRobustCommand{\regret}{\ensuremath{\mathrm{regret}}}
\DeclareRobustCommand{\idx}{\ensuremath{\mathrm{idx}}}

\renewcommand{\P}{\text{P}\xspace}
\newcommand{\NP}{\text{NP}\xspace}

\newcommand{\mdatspconstant}{\ensuremath{\alpha_{\text{MD}}}}
 
\title{Approximation Algorithms for Asymmetric Multi-Path TSP}

\ifspringer \author{
}
\authorrunning{
}
\institute{
}
\else 

\author{
Manuel Christalla\thanks{
Department of Computer Science, ETH Zurich, Switzerland.
Email: \href{mailto:mchristalla@ethz.ch}{mchristalla@ethz.ch}. 
}
\and
Vera Traub\thanks{
Department of Computer Science, ETH Zurich, Switzerland.
Email: \href{mailto:vtraub@ethz.ch}{vtraub@ethz.ch}.
}
}

\date{}
\fi

\begin{document}

\maketitle

\begin{abstract}
Asymmetric Multi-Path TSP is the generalization of the Asymmetric Traveling Salesperson Problem where we are looking for $k$ walks, each with prescribed start $s_i$ and prescribed end $t_i$, such that every vertex is visited by at least one of the walks. The objective is to minimize the total length of all walks.
In general, it is NP-hard to find any approximation for this problem in polynomial time (even if the approximation factor is allowed to depend on the number of vertices).
We show that one can obtain an $O(k)$-approximation in time $n^{O(k)}$, answering an open question by Friggstad [2013].

For the special case where all tour start vertices $s_i$ are identical and all end vertices~$t_i$ are identical, we obtain an $O(k)$-approximation in polynomial time, as well as an $(O(\frac{1}{\epsilon}), 1+\epsilon)$-bicriteria approximation. 
For the multi-depot setting, where $s_i =t_i$ for all $i$, we obtain a constant-factor approximation in polynomial time.

Finally, our results imply improved constant-factor approximation algorithms for Asymmetric Regret-Bounded Vehicle Routing and unit-weight Path ATSP.
\ifspringer

\else\fi
\end{abstract}

\section{Introduction}

Path TSP is the general version of the (metric) Traveling Salesperson Problem, where we are looking for a shortest tour from a given start vertex $s$ to a given end vertex~$t$ that visits all other vertices in between.
In the symmetric case, this problem was first studied by Hoogeveen \cite{hoogeeven91}, and a long line of work \cite{an15,sebö13,vygen16,gottschalk18,sebö19,traub19,zenklusen19} culminated in a  reduction from Path TSP to the classical TSP (i.e., the special case $s=t$) with an arbitrarily small loss in the approximation factor \cite{traub22path}.
In the asymmetric case, Feige and Singh \cite{feige07} gave an analogous reduction, which, however, loses a factor of $2+\epsilon$ in the approximation guarantee (for an arbitrary constant~$\epsilon >0$).
For ATSP, Vygen~\cite{vygen26} obtained an approximation factor $12 + 2\sqrt{2} + \epsilon \approx 14.83$ for general weights and $3+2\sqrt{2} + \epsilon \approx 5.83$ for unit weights.
For the path version, the techniques of \cite{vygen26,traub25} yield a $(15 + \epsilon)$-approximation.

In very recent work, Hajiaghayi and Yun~\cite{hajiaghayi26} report an improved approximation factor of $6+\sqrt{2}+\sqrt{15 + 6\sqrt{2}} + \epsilon \approx 12.2604$ for both ATSP and Path ATSP, and Li \cite{li26} reports a reduction from Path ATSP to ATSP with an arbitrarily small loss in the approximation factor.
The results of \cite{hajiaghayi26} and \cite{li26} were obtained in independent work parallel to ours.
Combining their techniques with ours might lead to further improvements of some of the constants in our results, which we do not yet consider in the current version of this paper.

\ifspringer\else\pagebreak\fi

In this work we focus on the natural multi-path generalization of ATSP and Path ATSP.

\begin{problem}[Asymmetric Multi-Path TSP]
An instance of Asymmetric Multi-Path TSP consists of a digraph $G=(V,E)$ with edge costs $c\colon E \to \bR_{\geq 0}$ and terminal pairs $(s_1,t_1), \dots, (s_k,t_k)$ with $s_i,t_i \in V$ for all $i \in [k]$.

The task is to find $s_i$-$t_i$-walks $P_i$ for each $i \in [k]$ such that $V = \bigcup_{i = 1}^k V(P_i)$ and the total cost $\sum_{i=1}^k c(P_i)$ is minimized, or to decide that no such collection of walks exists.
\end{problem}

For the symmetric special case (where $G$ is undirected) the currently best approximation factor is $1+\frac{2}{\sqrt{e}}\approx 2.2131$ due to Böhm, Friggstad, Mömke, and Spoerhase \cite{boehm25}. 
If the number $k$ of walks is a constant, then in the symmetric setting one can reduce the problem to TSP at an arbitrarily small loss in the approximation factor as shown in \cite{traub25}, building on \cite{traub22path}.
For the symmetric unit-weight case ($G$ undirected and $c\equiv 1$), Alimi, Dahlmeier, Mömke, Pabst, and Vargas Koch \cite{alimi26} recently gave a $2$-approximation, which was subsequently improved to a $1.9304$-approximation in \cite{dahlmeier26}.

In the general asymmetric setting, however, it is already NP-hard to decide if any feasible solution exists \cite{friggstad13}.
For the special case $k=2$, Friggstad \cite{friggstad13} gave an $O(\log n)$-approximation, where $n=|V|$.
Moreover, he asked whether one can obtain an approximation factor of $f(k) \cdot \log n$ in time $n^{g(k)}$ for some functions $f$ and $g$.
After Friggstad asked this question, the first constant-factor approximations for ATSP were achieved \cite{svensson20,traub22atsp} and thus it is now natural to ask for an $f(k)$-approximation in time $n^{g(k)}$.
We answer this question positively, even for $f$ and $g$ being linear~functions.

\begin{restatable}{theorem}{ThmConstantNumTerminalsMain}\label{thm:constant_num_terminals_main}
There is an algorithm that computes, for every Asymmetric Multi-Path TSP instance with $n$ vertices and $k$ terminal pairs, an $O(k)$-approximation in time $n^{O(k)}$.
\end{restatable}

Our proof of this result has two main ingredients.
First, we build on techniques from the existing constant-factor approximations for ATSP to obtain a bicriteria approximation for the special case where all start vertices $s_i$ are identical and all end vertices $t_i$ are identical.
Second, we provide a dynamic programming algorithm to reduce to this special case.
While this algorithm is a simple extension of the dynamic program from \cite{feige07}, our analysis is different.
For details, we refer to~\Cref{sec:constant_number_of_terminals}.

As deciding whether a feasible Multi-Path solution exists is NP-hard \cite{friggstad13}, we cannot hope for a polynomial-time approximation algorithm for arbitrary $k$, unless~$\P=\NP$.
Thus, in the following we consider special cases that admit polynomial-time approximation algorithms even when $k$ is arbitrarily large.

\subsection{The Single-Source/Single-Sink Case}
A natural special case of the Asymmetric Multi-Path TSP is when all  start vertices $s_i$ are identical and all end vertices $t_i$ are identical.

\begin{problem}[Asymmetric Single-Source/Single-Sink Multi-Path TSP]
An instance of Asymmetric Single-Source/Single-Sink Multi-Path TSP (Asymmetric SSSS Multi-Path TSP) consists of a digraph $G=(V,E)$ with edge costs $c\colon E \to \bR_{\geq 0}$ and two distinct vertices $s,t \in V$, as well as an integer $k \geq 1$.

The task is to find $s$-$t$-walks $P_1, \dots, P_k$ such that $V = \bigcup_{i = 1}^k V(P_i)$ and $\sum_{i=1}^k c(P_i)$ is minimized, or to decide that no such collection of walks exists.
\end{problem}
To decide feasibility of a given instance in polynomial time, we may contract all strongly connected components and then ask for an integral $s$-$t$-flow of value $k$ that covers every component.
We therefore focus on approximation guarantees and may always assume that instances are feasible.

A related variant, already studied by Friggstad \cite{friggstad13}, is the Asymmetric Multiple-Source/\allowbreak{}Multiple-Sink Multi-Path TSP, where the instance contains terminal sets $S=\parset{s_1,\dots,s_k},T=\parset{t_1,\dots,t_k} \subseteq V$ such that $S \cap T = \emptyset$ and $|S|=|T| = k \geq 1$.
The task is then to find a permutation~$\pi\colon [k] \to [k]$ and $s_i$-$t_{\pi(i)}$-walks $P_i$ for all $i \in [k]$ such that $V = \bigcup_{i = 1}^k V(P_i)$ and $\sum_{i=1}^k c(P_i)$ is minimized, or to decide that no such collection of walks exists.
However, from the perspective of approximation algorithms this problem is essentially equivalent to the Single-Source/Single-Sink variant (see~\Cref{sec:equivalence_to_other_variants}).
Thus, in the following we will focus on the Single-Source/Single-Sink case.

Friggstad \cite{friggstad13} gave an algorithm to compute a $(O(\frac{\log n}{\epsilon}), 1+\epsilon)$-bicriteria approximation in polynomial time.
For $\alpha, \beta \geq 1$, an \emph{$(\alpha,\beta)$-bicriteria approximation} for Asymmetric SSSS Multi-Path TSP is a solution consisting of at least~$k$ and at most~$\beta k$~walks, whose total cost is at most $\alpha$ times the cost of an optimal solution using exactly~$k$~walks.
We improve on Friggstad's bicriteria approximation by showing that the factor of $\log n$ in the approximation guarantee is not necessary.
Moreover, we show that in the unit-weight special case ($c\equiv 1$), we can even obtain a constant-factor approximation.

\begin{restatable}{theorem}{ThmSingleSourceSingleSinkMain}\label{thm:single_source_single_sink_main}
There is a polynomial-time algorithm that, given $\epsilon \in (0,1]$ and an instance of Asymmetric SSSS Multi-Path TSP, yields an $\left(O(\frac{1}{\epsilon}) , 1 + \epsilon \right)$-bicriteria approximation.

For the unit-weight special case of Asymmetric SSSS Multi-Path TSP, there is a $(5+2\sqrt{2} + \epsilon)$-approximation algorithm for any $\epsilon > 0$.
\end{restatable}

In particular, by choosing $\epsilon \coloneqq \tfrac{1}{k+1}$, \Cref{thm:single_source_single_sink_main} implies an $O(k)$-approximation algorithm for Asymmetric SSSS Multi-Path TSP.

Köhne, Traub, and Vygen \cite{köhne20} showed that any $\alpha$-approximation algorithm for unit-weight ATSP yields a $(2\alpha - 1)$-approximation algorithm for unit-weight Path ATSP.
Combining \cite{köhne20} with \cite{vygen26} gives an approximation guarantee of $5 + 4\sqrt{2} + \epsilon$ for unit-weight Path ATSP.
As this problem is just the special case of (unit-weight) Asymmetric SSSS Multi-Path TSP with $k=1$, \Cref{thm:single_source_single_sink_main} improves this approximation guarantee.
Since we analyze the cost of our solution relative to the LP optimum, this also implies an improved upper bound of $5+2\sqrt{2}$ on the integrality gap of the standard LP relaxation for unit-weight Path ATSP.
Remark~8.1 in the concurrent work~\cite{li26} suggests that their reduction might also apply to the unit-weight special case and thus imply an even better approximation factor of $3+2\sqrt{2}+\epsilon$ for unit-weight Path ATSP, but no improved upper bound on the integrality gap. However, to the best of our knowledge, this is not explicitly discussed in \cite{li26}.

We highlight that Friggstad's bicriteria approximation does not use an approximation algorithm for ATSP as a black box and we cannot simply combine his approach with the improved approximation algorithms for ATSP to prove the first part of \Cref{thm:single_source_single_sink_main}.
Moreover, directly applying existing techniques for Asymmetric Path TSP \cite{feige07, köhne20} would only lead to bicriteria approximations with a large constant-factor loss in the number of paths or $O(k^2)$-approximations for Asymmetric SSSS Multi-Path TSP. (See the discussion in \Cref{sec:bicriteria}.)

Our proof builds on techniques from \cite{köhne20}, but has two important new ingredients.
First, we show that the existing constant-factor approximation algorithms for ATSP can be modified to additionally guarantee the following.
For a given optimal LP solution $x$ and a given vertex $v$, we can ensure that the degree of $v$ in the computed tour is preserved, i.e., the out-degree of $v$ will be between $\lfloor x(\delta^+(v))\rfloor$ and $\lceil x(\delta^+(v))\rceil$, while the cost of the computed tour remains $O(c(x))$.
Second, to obtain our bicriteria approximation, we prove a key lemma about optimal dual solutions to the relaxed linear programming relaxation (where we allow $(1+\epsilon)k$ paths); see \Cref{lem:lp_path_scaling}.
Here, the idea is to view the optimal dual solution for the relaxed LP as a (feasible) dual solution to the original LP.
For details on the bicriteria approximation we refer to \Cref{sec:bicriteria}.

The second part of \Cref{thm:single_source_single_sink_main} will be proved in \Cref{sec:unweighted_ssss_multi-path_atsp}.

\subsection{Asymmetric Multi-Depot TSP} 
Another natural special case of the Asymmetric Multi-Path TSP is the multi-depot case, where we have $s_i=t_i$ for all $i\in [k]$.

\begin{problem}[Asymmetric Multi-Depot TSP]
An instance of Asymmetric Multi-Depot TSP consists of a digraph $G=(V,E)$ with edge costs $c\colon E \to \bR_{\geq 0}$ and a set $T = \{t_1, \dots, t_k \} \subseteq V$ of terminals.

The task is to find an Eulerian multi-subset $F\subseteq E$ of edges such that each strongly connected component of $(V,F)$ contains at least one terminal $t_i \in T$ and the total cost~$c(F)$ is minimized, or to decide that no such multi-set of edges exists.
\end{problem}

Note that there exists a feasible solution if and only if every strongly connected component of $G$ contains a terminal.
Hence, this can be decided in polynomial time, and we may therefore always assume that instances are feasible.

For the symmetric special case (where $G$ is undirected), a $(1+\frac{1}{\sqrt{2}})$-approximation was recently given by Zhao, Liu and Xiao \cite{zhao26}.
For the symmetric unit-weight case, Dahlmeier, Luqu\'{e}, Pabst, and Vargas Koch \cite{dahlmeier26} gave a $\frac{3}{2}$-approximation.

In the asymmetric setting, Friggstad \cite{friggstad13} obtained an $O(\log n)$-approximation algorithm by adapting the cycle cover algorithm of Frieze, Galbiati, and Maffioli \cite{frieze82} for ATSP.
We improve the approximation factor to a constant.

\begin{theorem}\label{thm:multi_depot_main}
There is a $(12 + 2\sqrt{2} + \epsilon)$-approximation algorithm for Asymmetric Multi-Depot TSP for any $\epsilon > 0$.

For the unit-weight special case of Asymmetric Multi-Depot TSP, there is a $(5 + 2\sqrt{2} + \epsilon)$-approximation algorithm for any $\epsilon > 0$.
\end{theorem}

Note that, for general weights, our approximation guarantee matches the $12 + 2\sqrt{2} + \epsilon$ bound  of \cite{vygen26} for ATSP.
\Cref{thm:multi_depot_main} will be proved in \Cref{sec:multi_depot_atsp}.

\subsection{Asymmetric Regret-Bounded Vehicle Routing}
Finally, our bicriteria approximation for Asymmetric SSSS Multi-Path TSP leads to an improved constant-factor approximation for the Asymmetric Regret-Bounded Vehicle Routing Problem.

\begin{problem}[Asymmetric Regret-Bounded Vehicle Routing Problem]
An instance of the Asymmetric Regret-Bounded Vehicle Routing Problem consists of a complete digraph~$G=(V,E)$ with a root vertex $r \in V$, edge costs $c\colon E \to \bR_{\geq 0}$ satisfying the triangle inequality, and a regret bound~$R \in \bR_{\geq 0}$.

The task is to find a minimum number of $r$-rooted paths $P_1, \dots, P_k$ such that every vertex is covered and each path $P_i$ has regret at most $R$, that is,
\[
\regret(P_i) \ \coloneqq\ c(P_i) \ -\ c(r,v_i) \ \leq\ R,
\]
where $v_i$ is the endpoint of $P_i$.
\end{problem}

For the symmetric variant of this problem (i.e., $c(v,w) = c(w,v)$ for all $v,w \in V$), the first constant-factor approximation algorithm was given by Friggstad and Swamy \cite{friggstad14}, and subsequently improved in \cite{friggstad17} and \cite{blauth23}, culminating in a $10$-approximation.
 
Friggstad and Swamy \cite{friggstad14} observed that any $(\alpha, \beta)$-bicriteria approximation algorithm for the Asymmetric SSSS Multi-Path TSP yields an $(\alpha + \beta)$-approximation algorithm for the Asymmetric Regret-Bounded Vehicle Routing Problem.
Combining this reduction with the results of \cite{friggstad13,friggstad13path} yields an $O(\log n)$-approximation.

We observe that any $\alpha$-approximation algorithm for ATSP directly yields a $(2\alpha + \epsilon,2\alpha)$-bicriteria approximation algorithm for Asymmetric SSSS Multi-Path TSP.
Via the reduction from~\cite{friggstad14} and the approximation algorithm of \cite{hajiaghayi26} for ATSP, this yields an approximation guarantee less than $50$ for Asymmetric Regret-Bounded Vehicle Routing.
Our bicriteria result for Asymmetric SSSS Multi-Path TSP implies an improvement:

\begin{restatable}{theorem}{ThmRegretBoundedVRP}\label{thm:regret_bounded_vrp}
There is a $23$-approximation algorithm for the Asymmetric Regret-Bounded Vehicle Routing Problem.
\end{restatable}

For details, we refer to \Cref{subsec:regret_bounded_vrp}.

\section{Approximating Asymmetric Multi-Path TSP with a Constant Number of Terminals}
\label{sec:constant_number_of_terminals}

In this section, we give the proof of \Cref{thm:constant_num_terminals_main}, which we restate here for convenience:

\ThmConstantNumTerminalsMain*

Consider an instance $\cI = (G,c, \{(s_i,t_i) \}_{i=1}^k)$ with $G=(V,E)$ and $n = |V|$, and let $S \coloneqq \parset{s_1, \dots, s_k}$ and $T \coloneqq \parset{t_1, \dots, t_k}$.
By contracting all strongly connected components of $G$ and using a straightforward dynamic program that processes components in the resulting topological order, we can decide in time $n^{O(k)}$ whether a feasible solution for $\cI$ exists.
We may therefore assume in the following that $\cI$ admits a feasible solution.
Moreover, by computing the metric closure, we assume without loss of generality that~$G$ is complete and $c$ satisfies the triangle inequality.
To make $G$ complete, we allow edges to have infinite cost.

The first step of our algorithm is to compute an initial solution of vertex-disjoint paths covering all non-terminal vertices, where the paths may start and end at arbitrary vertices.
We remark that this first step is similar to the approach used by Friggstad~\cite{friggstad13}.
\begin{restatable}{lemma}{LemInitialDpSolution}
\label{lem:initial_dp_solution}
We can compute in polynomial time an integer $\ell \leq 2k$ and vertex-disjoint paths $Q_1, \dots, Q_\ell$ covering $V \setminus (S \cup T)$ that satisfy
\begin{equation}\label{eq:initial_dp_solution}
\sum_{i=1}^\ell c(Q_i) \ \leq\ 51 \cdot \OPT_{\cI}
\end{equation}
where $\OPT_{\cI}$ is the cost of an optimal solution to the given instance~$\cI$.
\end{restatable}
\noindent\Cref{lem:initial_dp_solution} follows from our bicriteria approximation (\Cref{thm:single_source_single_sink_main} and \Cref{sec:equivalence_to_other_variants}).

For the remainder of this section, we fix paths $Q_1, \dots, Q_\ell$ as given by \Cref{lem:initial_dp_solution}.
In their reduction from Asymmetric Path TSP to Asymmetric TSP, Feige and Singh~\cite{feige07} also start with an initial collection of $s$-$t$-paths.
While their goal is to merge these paths into a single path, ours is to transform them into $k$ paths starting and ending at $s_i$ and $t_i$ for each $i \in [k]$, respectively.
(Observe that this is non-trivial even for $\ell=k$.)

Feige and Singh use these initial paths as follows.
Instead of considering arbitrary solutions, they only consider paths whose vertex order is compatible with the order along each of the initial paths.
We generalize this to the Multi-Path setting.

\begin{definition}[Precedence-Preserving Solution]\label{def:precedence_preserving}
A solution $P_1, \dots, P_k$ to the Asymmetric Multi-Path TSP instance $\cI$ is \emph{precedence-preserving} if there exists a linear order~$\preceq$ on $V \setminus (S \cup T)$ with the following property: For all $x,y \in V \setminus (S \cup T)$, if $x$ precedes $y$ on one of the paths $Q_1, \dots, Q_{\ell}, P_1, \dots, P_k$, then $x \preceq y$.
\end{definition}

Feige and Singh use a dynamic program to compute a cheapest precedence-preserving solution, which we generalize to our setting.

\begin{restatable}{lemma}{LemDP}\label{lem:dp}
Given $\ell \leq 2k$ vertex-disjoint paths $Q_1, \dots, Q_\ell$ covering $V \setminus (S \cup T)$, we can compute a cheapest precedence-preserving solution in time $n^{O(k)}$.
\end{restatable}

See \Cref{sec:DP} for a proof of \Cref{lem:dp}.
The key part of our analysis is to show that there exists a cheap precedence-preserving solution.
The idea is as follows.
We consider the auxiliary multi-graph containing all edges of the paths $Q_1, \dots, Q_\ell$ and all edges of an optimal solution to the instance $\cI$.
Observe that the optimal solution gives rise to $k$ paths visiting every strongly connected component of this multi-graph.
Within a single strongly connected component, any path $Q_j$ intersects this strongly connected component in an interval.
We will simply concatenate these (at most $\ell$) corresponding intervals, losing a factor~$O(k)$ in the approximation guarantee.
This leads to the following.

\begin{lemma}\label{lem:cheap_precedence_preserving}
There exists a precedence-preserving solution of cost at most $O(k) \cdot \OPT_{\cI}$.
\end{lemma}
\begin{proof}
Let $P^\ast_1, \dots,P^\ast_k$ be an optimal solution to the Asymmetric Multi-Path instance $\cI$.
By the triangle inequality, we may assume that each vertex of $V \setminus (S \cup T)$ appears on at most one of these walks, and no terminal vertex of~$S \cup T$ appears as an internal vertex.
Let~$H$ be the multi-graph with vertex set $V \setminus (S \cup T)$ obtained by taking the union of the edges of $P^\ast_1, \dots,P^\ast_k$ and $Q_1, \dots, Q_\ell$, excluding those that are incident to vertices of~$S \cup T$.
Let $\mathcal{K}$ be the collection of strongly connected components of $H$.

Fix a component $C \in \mathcal{K}$.
For every $j\in[\ell]$, the vertices $V(Q_j)\cap V(C)$ form a (possibly empty) contiguous subpath of $Q_j$, because otherwise $C$ would not be a maximal strongly connected subgraph of $H$.
Among these subpaths $Q_1[V(C)],\dots,Q_\ell[V(C)]$
concatenate the nonempty ones in an arbitrary order, and denote the resulting path by~$R_C$.
\ifspringer\begin{samepage}\else\fi Note that $V(R_C) = V(C)$.
Because $C$ is strongly connected, the cost of any edge connecting two subpaths is at most $c(E(C))$ by the triangle inequality.
Hence,
\begin{equation}\label{eq:ssc_bound}
c(R_C) \ \leq \ (\ell - 1) \cdot c(E(C)) \ +\ \sum_{j=1}^\ell c\big( Q_j[V(C)] \big) \ \leq \ \ell \cdot c(E(C)).
\end{equation}

For every $C\in \mathcal{K}$, choose $\idx(C)\in[k]$ such that $V(P^\ast_{\idx(C)}) \cap V(C) \neq \emptyset$.
\ifspringer\end{samepage}\else\fi
We construct $P_i$ as the path that starts in $s_i$, then follows all paths $R_C$ with $\idx(C)=i$ in increasing topological order of the corresponding strongly connected components $C$, and ends in $t_i$.

We now bound the cost of the paths $P_i$.
For each component $C \in \mathcal{K}$ assigned to $P_i$, let $a_C$ and $b_C$ be the first and last vertices of $P_{i}^\ast$ contained in $C$.
Follow $P^\ast_i$ between the assigned components, and within each such component $C$, connect $a_C$ to the start of~$R_C$, traverse~$R_C$, and connect its end to $b_C$.
Because each $C \in \mathcal{K}$ is strongly connected, by the triangle inequality we obtain
\begin{equation}\label{eq:path_bound}
c(P_i) \ \leq\ c(P_i^\ast) \ +\ \sum_{\substack{C\in \mathcal{K}:\\ \idx(C)=i}} \Big( c(R_C) + 2c(E(C)) \Big).
\end{equation}
Combining \eqref{eq:ssc_bound}\ and \eqref{eq:path_bound} yields
\begin{align*}
\sum_{i=1}^k c(P_i)
\ &\leq\ (\ell+2) \cdot \sum_{C\in \mathcal{K}} c(E(C))\ +\ \OPT_{\cI}\\
\ &\leq\ (\ell+2) \cdot \left(
   \sum_{j=1}^{\ell}  c(Q_j) + \sum_{i=1}^k c(P^\ast_i) 
   \right) \ +\ \OPT_{\cI}\\
\ \overset{\eqref{eq:initial_dp_solution}}&{\leq} \ (52\ell + 105) \cdot \OPT_{\cI} \\
\ &\leq \ O(k) \cdot \OPT_{\cI},
\end{align*}
since $\ell \leq 2k$.

Finally, order the vertices of $V \setminus (S \cup T)$ first by the topological order of their component $C\in \mathcal{K}$ and within each component $C$ by their order on $R_C$.
This yields a linear order that is consistent with every $Q_j$ and every $P_i$.
Therefore, the constructed solution $P_1, \dots, P_k$ is precedence-preserving.
\end{proof}
Finally, \Cref{lem:initial_dp_solution}, \Cref{lem:dp} and \Cref{lem:cheap_precedence_preserving} together imply \Cref{thm:constant_num_terminals_main}. 

\section{A Bicriteria Approximation for Asymmetric Single-Source/\\Single-Sink Multi-Path TSP}
\label{sec:bicriteria}

In this section, we prove the first part of \Cref{thm:single_source_single_sink_main}.
That is, we provide an $\big(O(\frac{1}{\epsilon}), 1+ \epsilon\big)$-bicriteria approximation algorithm for Asymmetric SSSS Multi-Path TSP.
To this end, we consider the following natural LP relaxation for a given instance $(G,c,s,t,k)$:

\begin{align}
&\text{min} \ & c(x) \notag \\ 
&\text{s.t.} & x(\delta^-(v)) - x(\delta^+(v)) &\ =\ 
\begin{cases} 
-k, \quad&\text{if $v=s$}\\ 
k, \quad&\text{if $v=t$} \\ 
0, \quad&\text{if $v \in V \setminus \{s,t\}$}
\end{cases} \quad && \tag{SSSS-P}\label{lp:multi_path_primal} \\ 
&& x(\delta(U)) &\ \geq\ 2 \quad && (\emptyset \neq U \subseteq V \setminus \parset{s,t}) \notag \\ 
&& x_e &\ \geq\ 0 \quad && (e \in E). \notag
\end{align}
In particular, its dual is given by
\begin{align}
&\text{max} \ & k (a_t - a_s) \ +\ \sum_{\emptyset \neq U \subseteq V \setminus \parset{s,t}} 2y_U \tag{SSSS-D}\label{lp:multi_path_dual} \\
&\text{s.t.} & a_w - a_v + \sum_{U\colon e \in \delta(U)} y_U &\ \leq\ c(e) && (e=(v,w) \in E) \notag \\
&& y_U &\ \geq\ 0 && (\emptyset \neq U \subseteq V \setminus \parset{s,t}). \notag
\end{align}

In prior work on Path ATSP \cite{feige07,köhne20} the common approach is to add an artificial backward edge $(t,s)$ (or some other path from $t$ to $s$) and then apply a constant-factor approximation to the resulting instance of ATSP.
In a second step one removes all copies of the edge $(t,s)$ from the tour, which results in a collection of $s$-$t$ walks.
However, this approach does not provide any control of the number of resulting $s$-$t$ walks.
If we have too few walks, this can be easily addressed by adding a sufficient number of shortest $s$-$t$ paths.
If we have too many walks, the situation is more tricky and some of the $s$-$t$ walks need to be merged into a single one.
To account for the cost of this merging step, the cost of the artificial backward edge $(t,s)$ is typically chosen as the merging cost required to reduce the number of $s$-$t$ walks by one.

However, in our multi-path setting merging becomes more difficult and it is not clear how to reduce the number of paths by one at a cost of less than some constant factor times $k\cdot \LP$ or $k\cdot \OPT$ (and already this is not obvious).

Moreover, in order to turn an optimal solution to \eqref{lp:multi_path_primal} into a feasible solution to the ATSP~LP, we would need to give the backward edge $(t,s)$ an LP value of $k$. (The same applies when working with integral solutions instead of LP solutions.)
Overall, this would increase the cost by a factor proportional to $k^2$.

\paragraph{Degree-preserving approximations.}
Our first key idea to avoid this loss is to adapt the existing constant-factor approximation algorithm for ATSP to guarantee an additional degree-preserving property.
We will prove this in a more general form for the Multi-Depot ATSP, which will also imply \Cref{thm:multi_depot_main}.
To provide a precise description of the degree-preserving guarantee, consider the following LP relaxation for Multi-Depot ATSP:
\begin{equation}
\tag{MD-P}\label{lp:multi_depot_primal}
\begin{aligned}
&\text{min} \ & c(x) \\ 
&\text{s.t.} & x(\delta^-(v)) &\ =\ x(\delta^+(v)) \quad && (v \in V) \\ 
&& x(\delta(U)) &\ \geq\ 2 \quad && (\emptyset \neq U \subseteq V \setminus T) \\ 
&& x_e &\ \geq\ 0 \quad && (e \in E).
\end{aligned}
\end{equation}

Then we define degree-preserving solutions as those that preserve the degree of the depot vertices with respect to a given LP solution.

\begin{definition}[Degree-preserving]
Let $(G,T,c)$ be an instance of Asymmetric Multi-Depot TSP.
A feasible solution $F$ is \emph{degree-preserving} w.r.t.\ a feasible solution $x$ to \eqref{lp:multi_depot_primal} if
\begin{equation*}
\floor{x(\delta^-(t))} \ \leq\ |\delta^-_F (t)| \ \leq\ \ceil{x(\delta^-(t))}
\end{equation*}
for all $t \in T$.
\end{definition}

We adapt existing algorithms for ATSP to prove that we can achieve a constant-factor approximation that is additionally degree-preserving.
Throughout the remainder of this section, we state all subsequent bounds in terms of the approximation factor $\mdatspconstant$ of our degree-preserving approximation algorithm for Multi-Depot ATSP to make the dependence on this guarantee explicit.
In \Cref{sec:multi_depot_atsp}, we prove that $\mdatspconstant \leq 12 + 2\sqrt{2} + \epsilon$ by proving the following theorem.

\begin{restatable}{theorem}{ThmWeightedMultiDepotATSP}\label{thm:weighted_multi_depot_atsp}
Let $\mdatspconstant > 12 + 2\sqrt{2} \approx 14.83$.
Given an instance  $(G,T,c)$ of Asymmetric Multi-Depot TSP and an optimal solution $x$ to \eqref{lp:multi_depot_primal}, we can compute a degree-preserving solution $F \subseteq \supp(x)$ of cost at most $ \mdatspconstant \cdot c(x)$ in polynomial time.

In the unit-weight case ($c \equiv 1$), it suffices to choose $\mdatspconstant > 5+2\sqrt2 \approx 7.83$ and $x$ to be a feasible solution to \eqref{lp:multi_depot_primal}.
\end{restatable}

In order to use \Cref{thm:weighted_multi_depot_atsp} for our bicriteria approximation for the Asymmetric SSSS Multi-Path TSP, we will apply it in the special case $|T|=1$ with a single depot vertex.
We will construct an ATSP instance by adding a new depot vertex $v^*$ and a backward path consisting of edges $(t,v^*)$ and $(v^*,s)$. 
The degree-preserving property allows us to control how often this backward path is used.
This avoids the necessity of merging~$s$-$t$~walks.
We choose the total cost of this backward path as $a_s - a_t$, so that sending $k$ units of flow along it extends $x$ to an \emph{optimal} solution of the multi-depot LP, as required by \Cref{thm:weighted_multi_depot_atsp}.
Since this cost can be negative, we use the dual potentials to obtain nonnegative edge costs while preserving the cost of the circulation.
Via this approach, we obtain an algorithm that returns a solution whose cost can be bounded by the LP value plus some constant factor times the second summand of the objective in~\eqref{lp:multi_path_dual}.

\begin{lemma}
\label{lem:dual_relative_algo}
Given an instance $\cI = (G,c,s,t,k)$ of Asymmetric SSSS Multi-Path TSP and optimal solutions $x$ and $(a,y)$ to \eqref{lp:multi_path_primal} and \eqref{lp:multi_path_dual}, respectively, we can compute a feasible solution to $\cI$ of cost at most $k(a_t - a_s) + \mdatspconstant \cdot \sum_{\emptyset \neq U \subseteq V \setminus \parset{s,t}} 2y_U$ in polynomial time.
\end{lemma}
\begin{proof}
Let $G=(V,E)$.
We construct an auxiliary graph $G^\prime = (V^\prime, E^\prime)$ by adding a new vertex $v^\ast$ to $V$ together with edges $(t,v^\ast)$ and $(v^\ast, s)$.
We then extend $x$ to $x^\prime$ by assigning both newly added edges the value $k$.
Consider the cost function
\[
c^\prime(v,w) \ \coloneqq\
\begin{cases}
c(v,w) + a_v - a_w \quad&\text{if } (v,w) \in E,\\
0 & \text{otherwise}.
\end{cases}
\]
Since $(a,y)$ is feasible for \eqref{lp:multi_path_dual}, we have $c^\prime(e) \geq 0$ for all $e \in E^\prime$.
Observe that $x^\prime$ and $(0,y)$ are feasible solutions to \eqref{lp:multi_depot_primal} and \eqref{lp:multi_depot_dual} for the Asymmetric Multi-Depot TSP instance $(G^\prime, T^\prime, c^\prime)$, respectively, where we choose the depot set $T^\prime = \{v^\ast\}$.

Using the fact that $x$ is an $s$-$t$-flow of value $k$ and strong duality for $x$ and $(a,y)$, we have
\[
c^\prime(x^\prime) 
\ =\ c(x) \ + \ k(a_s - a_t)
\ =\ \sum_{\emptyset \neq U \subseteq V \setminus \parset{s,t}} 2y_U.
\]
This implies, again by strong duality, that $x^\prime$ and $(0,y)$ are optimal primal-dual solutions to \eqref{lp:multi_depot_primal} and \eqref{lp:multi_depot_dual}, respectively.

We apply \Cref{thm:weighted_multi_depot_atsp} to the instance $(G^\prime, T^\prime, c^\prime)$ and fractional solution $x^\prime$ to obtain a degree-preserving solution $F^\prime \subseteq \supp(x^\prime)$.
In particular, we have $|\delta^+_{F^\prime} (v^\ast)| = |\delta^-_{F^\prime} (v^\ast)| = x^\prime(\delta^-(v^\ast)) = k$.
Let $F$ be the multi-set of edges in $G$ obtained from $F^\prime$ by removing all copies of the edges $(t,v^\ast)$ and $(v^\ast, s)$.
Because $v^\ast$ has in- and out-degree exactly $k$ with respect to $F^\prime$, this allows us to decompose $F$ into $k$~walks from $s$ to $t$, denoted by $P_1, \dots, P_k$, which form a feasible SSSS Multi-Path solution.
\begin{align*}
\sum_{i=1}^k c(P_i)
\ =\ c^{\prime}(F^\prime) \ +\ k(a_t - a_s) 
\ &\leq\ \mdatspconstant \cdot c^\prime(x^\prime) \ +\ k(a_t - a_s) \\
&= k(a_t - a_s) \ +\ \mdatspconstant \cdot \sum_{\emptyset \neq U \subseteq V \setminus \parset{s,t}} 2y_U .
\end{align*}
\end{proof}

Note that \Cref{lem:dual_relative_algo} yields a solution of cost at most $\LP + (\mdatspconstant - 1) \cdot \sum_{\emptyset \neq U \subseteq V \setminus \parset{s,t}} 2y_U$, where $\LP$ denotes the optimal objective value of \eqref{lp:multi_path_dual}.
To obtain a bound with respect to the LP value, we would like to bound the second term in terms of $\LP$.
By generalizing a proof from \cite{köhne20}, we obtain the following bound.

\begin{restatable}{lemma}{LemWorstCaseGap}\label{lem:worst_case_gap}
Let $(G,c,s,t,k)$ be an instance of Asymmetric SSSS Multi-Path TSP, where $G=(V,E)$ is the support graph of an optimal primal LP solution to \eqref{lp:multi_path_primal}.
Then we can in polynomial time compute an optimal dual solution $(a,y)$ to \eqref{lp:multi_path_dual} such that
\begin{equation}
\label{eq:worst_case_gap}
\sum_{\emptyset \neq U \subseteq V \setminus \parset{s,t}} 2y_U \ \leq\ (k+1) \cdot \LP,
\end{equation}
where $\LP$ denotes the optimal objective value of \eqref{lp:multi_path_dual}.
\end{restatable}

See \Cref{sec:proof_worst_case_gap} for the proof details.
We remark that the bound~\eqref{eq:worst_case_gap} is tight; see \Cref{fig:worst_case_gap_tight}.
Combining \Cref{lem:dual_relative_algo} and \Cref{lem:worst_case_gap} yields an $O(k)$-approximation for the Asymmetric SSSS Multi-Path TSP.

\paragraph{Obtaining the bicriteria guarantee.}
In order to obtain the claimed bicriteria approximation for a fixed $\epsilon >0$, we consider a solution to the dual program \eqref{lp:multi_path_dual} for $k' \coloneqq \lfloor (1+\epsilon) k\rfloor$.
A useful observation is that any dual LP solution for $k'$ paths is also a feasible dual solution for $k$ paths.
This leads to the following key lemma, which allows us to bound the sum of the $y'$-variables (for the dual solution for $k'$ paths) essentially by~$O(\frac{1}{\epsilon})$ times the LP value for $k$ paths.

\begin{lemma}
\label{lem:lp_path_scaling}
Let $(G,c,s,t,k)$ with $G = (V,E)$ be an instance of Asymmetric SSSS Multi-Path TSP and $\epsilon > 1/k$.
Denote by $\LP$ the value of \eqref{lp:multi_path_primal} for the instance $(G,c,s,t,k)$.
Moreover, let $(a^\prime,y^\prime)$ be an optimal solution to \eqref{lp:multi_path_dual} for $(G,c,s,t,k^\prime)$, where $k^\prime \coloneqq \floor{(1+\epsilon)k}$.
Then we have
\begin{equation}
\label{eq:lp_path_scaling}
\sum_{\emptyset \neq U \subseteq V \setminus \parset{s,t} } 2y^\prime_U 
\ \leq\ \frac{(1 + \epsilon)k}{\epsilon k - 1} \cdot \LP.
\end{equation}
\end{lemma}
\begin{proof}
Let $\LP^\prime$ denote the value of \eqref{lp:multi_path_primal} for the instance $(G,c,s,t,k^\prime)$.
Moreover, let $A^\prime \coloneqq a^\prime_t - a^\prime_s$ and $Y^\prime \coloneqq \sum_{\emptyset \neq U \subseteq V \setminus \parset{s,t}} y^\prime_U$.
In particular, $\LP^\prime = k^\prime A^\prime + 2Y^\prime$.
Because $(a^\prime,y^\prime)$ is a feasible solution for the linear program \eqref{lp:multi_path_dual} for the instance $(G,c,s,t,k)$, we have
$
\LP 
 \geq  k A^\prime  + 2Y^\prime
$.

If $A^\prime \geq 0$, then \eqref{eq:lp_path_scaling} is already satisfied.
Hence, we may assume that $A^\prime \leq 0$.
We have  $k^\prime = \floor{(1+\epsilon) k} \geq (1+\epsilon)k - 1$, so $\frac{k^\prime + 1}{1+ \epsilon} \geq k$.
Therefore, $\LP$ is at least
\begin{equation*}
\begin{aligned}
k A^\prime \ +\ 2Y^\prime
\ &\geq\ \frac{k^\prime + 1}{(1+\epsilon)k^\prime} \cdot k^\prime A^\prime  \ + \ 2Y^\prime \\
\ &=\ \frac{k^\prime + 1}{(1+\epsilon)k^\prime} \cdot (\LP^\prime - 2Y^\prime  )  \ + \ 2Y^\prime \\
\ &\geq \ \left( 1 - \frac{k^\prime + 1}{(1+\epsilon)k^\prime} \right) \cdot 2Y^\prime. 
\end{aligned}
\end{equation*}
Note that
\begin{equation*}
1 - \frac{k^\prime + 1}{(1+\epsilon)k^\prime}
\ = \ \frac{\epsilon k^\prime - 1}{(1+\epsilon)k^\prime}
\ = \ \frac{\epsilon - \frac{1}{k^\prime}}{1+\epsilon}
\ \geq \ \frac{\epsilon - \frac{1}{k}}{1+\epsilon}
\ = \ \frac{\epsilon k - 1}{(1+\epsilon)k}.
\end{equation*}
\end{proof}

Combining \Cref{lem:dual_relative_algo}, applied to the instance $(G,c,s,t,k')$, with \Cref{lem:lp_path_scaling}, and using the fact that $\LP' \leq (1+\epsilon)\cdot \LP$, yields the desired bicriteria approximation (first part of \Cref{thm:single_source_single_sink_main}).

\begin{theorem}
There is a polynomial-time algorithm that, given $\epsilon \in (0,1]$ and an instance of Asymmetric SSSS Multi-Path TSP, yields an $\big(O(\frac{1}{\epsilon}) , 1 + \epsilon \big)$-bicriteria approximation.
\end{theorem}
\begin{proof}
Let $0 < \epsilon \leq 1$ and let $\cI \coloneqq (G,c,s,t,k)$ be an instance of Asymmetric SSSS Multi-Path TSP.
We may assume without loss of generality that $G=(V,E)$ is the support graph of an optimal primal LP solution to \eqref{lp:multi_path_primal}, since removing edges outside of the support does not increase the LP value.

Use \Cref{lem:worst_case_gap} to compute an optimal dual LP solution $(a,y)$ to \eqref{lp:multi_path_dual} that satisfies $\sum_{\emptyset \neq U \subseteq V \setminus \parset{s,t}} 2y_U \leq (k+1) \cdot \LP$.
We define $\tau \coloneqq 1 + \sqrt{2}$ and distinguish two cases:

First, suppose that $\epsilon \leq \frac{\tau}{k}$.
Then, by \Cref{lem:dual_relative_algo}, we compute a feasible solution to $\cI$ of cost at most
\begin{align*}
k(a_t - a_s) \ +\ \mdatspconstant \cdot \sum_{\emptyset \neq U \subseteq V \setminus \{s,t\}} 2y_U
\ \overset{\eqref{eq:worst_case_gap}}&{\leq}\ \LP \ +\ (\mdatspconstant - 1) \cdot (k+1) \cdot \LP \\
\ &\leq \ \LP \cdot \Big( \mdatspconstant \cdot \frac{\tau}{\epsilon} + \mdatspconstant \Big) \\
\ &\leq \ \frac{\mdatspconstant \cdot (2 + \sqrt{2})}{\epsilon} \cdot \LP.
\end{align*}

Second, suppose that $\epsilon > \frac{\tau}{k}$.
In particular, $\epsilon > \frac{1}{k}$.
Let $k^\prime \coloneqq \floor{(1+\epsilon)k}$ and compute an optimal dual solution $(a^\prime,y^\prime)$ to \eqref{lp:multi_path_dual} for the instance $\cI^\prime \coloneqq (G,c,s,t,k^\prime)$ and denote by $\LP^\prime$ its value.
Note that
\begin{equation}
\label{eq:scale_flow_values}
\LP^\prime \ \leq\ (1+\epsilon) \LP \ \leq\ 2 \LP,
\end{equation}
since we can scale up all primal variable values.
Moreover,
\begin{equation}
\label{eq:epsk_bound}
\epsilon k -1 
\ =\ \left(1- \frac{1}{\tau}\right)\epsilon k \ +\ \frac{\epsilon k}{\tau} \ -\ 1
\ > \ \left(1- \frac{1}{\tau}\right)\epsilon k
\ = \ (2- \sqrt{2})\cdot \epsilon k.
\end{equation}
By \Cref{lem:dual_relative_algo} together with \Cref{lem:lp_path_scaling}, we obtain a feasible solution to $\cI^\prime$ (consisting of $k^\prime \leq (1+\epsilon)k$ walks) of cost at most
\begin{align*}
k^\prime(a^\prime_t - a^\prime_s) \ +\ \mdatspconstant \cdot \sum_{\emptyset \neq U \subseteq V \setminus \{s,t\}} 2y^\prime_U
\ \overset{\eqref{eq:lp_path_scaling}}&{\leq} \ \LP^\prime \ +\ (\mdatspconstant - 1) \cdot \frac{(1 + \epsilon)k}{\epsilon k - 1} \cdot \LP \  \\
\ \overset{\eqref{eq:scale_flow_values}}&{\leq} \ 2\LP \ +\ (\mdatspconstant-1) \cdot \frac{2k}{\epsilon k - 1} \cdot \LP \\
\ \overset{\eqref{eq:epsk_bound}}&{\leq} \ 2\LP \ +\ (\mdatspconstant-1) \cdot \frac{2 + \sqrt{2}}{\epsilon} \cdot \LP \\
\ &\leq \ \frac{\mdatspconstant \cdot (2 + \sqrt{2})}{\epsilon} \cdot \LP.
\end{align*}
Hence, in both cases, we obtain a feasible solution that uses between $k$ and $(1+\epsilon)k$ walks and has cost at most $\frac{\mdatspconstant \cdot (2 + \sqrt{2})}{\epsilon}$ times the optimum value of \eqref{lp:multi_path_primal}.
By \Cref{thm:weighted_multi_depot_atsp}, this approximation factor is at most $\frac{51}{\epsilon}$.
\end{proof}

\section{Open Questions}

For the Asymmetric Multi-Path TSP, we showed that there is an $f(k)$-approximation algorithm with running time $n^{g(k)}$ for $f = O(k)$ and $g = O(k)$, where $n$ is the number of vertices and $k$ is the number of terminal pairs (\Cref{thm:constant_num_terminals_main}).
Since determining whether a feasible solution exists is NP-hard for this problem \cite{friggstad13}, we cannot hope for such an algorithm with running time polynomial in $n$ and $k$, unless $\P = \NP$.
However, it remains open whether the approximation guarantee can be improved and made independent of~$k$.

\begin{openquestion}
Is there a $\rho$-approximation algorithm for the Asymmetric Multi-Path TSP in time $n^{g(k)}$ for some function $g$ and some constant $\rho \geq 1$, which is independent of the number $k$ of terminal pairs?
\end{openquestion}

Moreover, it remains open whether a true constant-factor approximation for the Single-Source/\allowbreak{}Single-Sink Multi-Path TSP exists.
Our algorithm substantially exploits the standard LP relaxation~\eqref{lp:multi_path_primal}.
An upper bound on its integrality gap of $O(k)$ follows from \Cref{lem:dual_relative_algo,lem:worst_case_gap}. (Here $k$ is again the number of walks from the source to the sink.)

\begin{openquestion}
Is the integrality gap of the standard LP relaxation~\eqref{lp:multi_path_primal} for the Single-Source/Single-Sink Multi-Path TSP bounded by a constant?
\end{openquestion}

 \section*{Acknowledgements}
The authors used generative AI tools (ChatGPT) for language editing to improve grammar, wording, and readability. 
All new ideas, proofs, and results presented in this paper were developed by the authors, who take full responsibility for its content.

\printbibliography

\appendix
\crefalias{section}{appendix}
\crefalias{subsection}{appendix}

\section{Dynamic Program for Precedence-Preserving Solutions}
\label{sec:DP}

In this section, we prove \Cref{lem:dp}, which we restate here for convenience:

\LemDP*

\begin{proof}
We generalize the dynamic program of Feige and Singh~\cite{feige07} to our setting.
The dynamic program constructs the paths $P_1, \dots, P_k$ simultaneously.
A state consists of
\begin{itemize}
\item
for every $j \in [\ell]$, a prefix $R_j$ of $Q_j$, and

\item
for every $i \in [k]$, the current last vertex of $P_i$.
\end{itemize}
The dynamic program maintains the invariant that the set of vertices covered by the partial paths~$P_i$ is precisely $S \cup \bigcup_{j=1}^\ell V(R_j)$.
For every state, we compute a minimum-cost collection of partial paths realizing this state.
Initially, we consider the state where all prefixes $R_j$ are empty and each $P_i$ consists of just the single vertex $s_i$.

To transition to a successor state, we choose indices $i \in [k]$ and $j \in [\ell]$ with $R_j \neq Q_j$.
Let $v$ be the first vertex of $Q_j$ not contained in $R_j$.
We append $v$ to $P_i$ and extend $R_j$ by $v$, updating the cost accordingly.
We process all states in increasing order of $\sum_{j=1}^{\ell} |V(R_j)|$.

After all states have been processed, we enumerate all final states (where $R_j = Q_j$ for all $j$).
If $v_i$ is the current last vertex of $P_i$ in this state, we append the edge $(v_i,t_i)$ for every $i\in[k]$.
Among all solutions we obtain from the final states, we return a cheapest one.

The order in which we added vertices to obtain the returned solution is a linear order that is consistent with each $Q_j$ and each constructed path $P_i$.
Hence, we indeed obtain a precedence-preserving solution.
Conversely, consider any precedence-preserving solution and its witnessing linear order.
By the triangle inequality, we may assume that every non-terminal vertex appears at most once among the walks, and that no terminal appears as an internal vertex.
Consider the vertices according to the linear order and append each vertex to the path $P_i$ containing it in the given solution.
Whenever a vertex of $Q_j$ is considered, the consistency of the linear order with $Q_j$ implies that it is the first vertex following the already considered prefix $R_j$ of $Q_j$.
Hence, a straightforward induction shows that the dynamic program constructs a solution which is not more expensive than this solution.

Finally, there are $n^{O(k +\ell)}$ different states and each state has $O(k\ell)$ successor states.
Because $\ell \leq 2k$, the overall running time is $n^{O(k)}$.
\end{proof} 

\section{Obtaining Degree-Preserving Solutions for Asymmetric Multi-Depot TSP}
\label{sec:multi_depot_atsp}

In this section, we revisit and adapt the ATSP approximation framework of \cite{traub25,vygen26} to the multi-depot version to prove \Cref{thm:weighted_multi_depot_atsp}, which directly implies \Cref{thm:multi_depot_main}.
Most of this framework carries over unchanged, and we only need to address two parts.
First, we use the existing notion of \emph{vertebrate pairs} and reduce the multi-depot variant to such instances by adding an artificial cycle through all depot terminals as a \emph{backbone}.
Second, we observe that a degree-preserving variant of \emph{subtour covers} can be used within Svensson's algorithm~\cite{svensson15} to give us the desired degree guarantees on the depot terminals.

\subsection{Strongly Laminar Instances}

The dual of \eqref{lp:multi_depot_primal} is
\begin{equation}
\tag{MD-D}\label{lp:multi_depot_dual}
\begin{aligned}
&\text{max} \ & \sum_{\emptyset \neq U \subseteq V \setminus T} 2y_U \\
&\text{s.t.} & a_w - a_v + \sum_{U\colon e \in \delta(U)} y_U &\ \leq\ c(e) \quad && (e=(v,w) \in E) \\
&& y_U &\ \geq\ 0 \quad && (\emptyset \neq U \subseteq V \setminus T).
\end{aligned}
\end{equation}

The first step in the constant-factor approximation for ATSP is to reduce general instances to \emph{strongly laminar instances}, which can be analogously defined in our setting.

\begin{definition}[Strongly laminar instance]
A \emph{strongly laminar instance} $\cI$ of Asymmetric Multi-Depot TSP is a tuple $(G,T,\cL,x,y)$, where
\begin{itemize}
\item
$G=(V,E)$ is a digraph,

\item
$T \subseteq V$ is a set of terminals,

\item
$\cL$ is a laminar family of subsets of $V \setminus T$ such that $G[U]$ is strongly connected for all $U \in \cL$,

\item
$x$ is feasible for \eqref{lp:multi_depot_primal}, $x_e > 0$ for all $e \in E$, and $x(\delta(U)) = 2$ for all $U \in \cL$, and

\item
$y\colon \cL \to \bR_{\geq 0}$.
\end{itemize}
Its induced cost function is
\[
c^y(e) \ \coloneqq\ \sum_{U \in \cL\colon e \in \delta(U)} y_U,
\]
and we write
\[
\LP(\cI) \ \coloneqq\ c^y(x) \ =\ \sum_{U \in \cL} 2y_U.
\]
\end{definition}

To reduce to such instances, we use the given optimal primal solution $x$ to \eqref{lp:multi_depot_primal}, and compute an optimal dual solution $(a,y)$ to \eqref{lp:multi_depot_dual}.
By standard uncrossing, we may assume that the support $\cL \coloneqq \{ U \colon y_U > 0 \}$ of $y$ is laminar.
Moreover, as in the ATSP case \cite{traub25}, the dual solution can be chosen such that $G_x[U]$ is strongly connected for every $U \in \cL$, where $G_x$ is the support graph of $x$.
By complementary slackness, 
$x(\delta(U)) = 2$ for all $U \in \cL$ and $c(e) = c^y(e) + a_w - a_v$ for all $e=(v,w) \in E(G_x)$.
Since $x$ is a circulation, $c^y(x) = c(x)$.
Likewise, for any Eulerian multi-set $F \subseteq E(G_x)$, we have $c^y(F) = c(F)$.
This yields the following lemma.

\begin{lemma}\label{lem:reduction_to_strongly_laminar}
Let $\alpha \geq 1$.
If there is a polynomial-time algorithm that, for every strongly laminar instance $\cI ^\prime \coloneqq (G^\prime,T^\prime,\cL^\prime,x^\prime,y^\prime)$, computes a degree-preserving solution of (induced) cost at most~$\alpha \cdot\LP(\cI^\prime)$,
then there is a polynomial-time algorithm that, given any instance $(G,T,c)$ of Asymmetric Multi-Depot TSP and an optimal solution $x$ to \eqref{lp:multi_depot_primal}, computes a degree-preserving feasible solution $F \subseteq \supp(x)$ of cost $c(F) \leq \alpha \cdot c(x)$. 
\end{lemma}
A detailed proof is given in \cite{traub25} for standard ATSP. 
However, the same proof carries over directly to our setting.

For unit-weight Asymmetric Multi-Depot TSP, we slightly relax the definition of strongly laminar instances by allowing
\[ 
\cL \ \coloneqq\ \{\{ v \}\colon v \in V \}, \qquad y_v \ \coloneqq\ y_{\{v\}} \ =\ 1/2,
\]
even though this might violate the conditions $x(\delta(U)) = 2$ and $U \subseteq V \setminus T$ for $U \in \cL$.
In this case, we define $\LP(\cI) \coloneqq c^y(x)$.
This will allow us to treat both the unit-weight case and the general case simultaneously.

\subsection{Vertebrate Pairs}

The next step is to reduce the problem to \emph{vertebrate pairs}.

In the constant-factor approximation for Asymmetric TSP, a strongly laminar instance is solved by first contracting each maximal non-singleton set $L \in \cL$ to a single vertex.
The resulting contracted instance is node-weighted and can be solved by Svensson's algorithm \cite{svensson15, traub25}.
All contracted sets are then uncontracted and handled recursively.

However, a direct application of this approach does not prevent the same set $L \in \cL$ from being charged at multiple levels of the recursion.
To avoid this, the algorithm further adds a connected Eulerian multi-graph, called the \emph{backbone}.
Laminar sets visited by the backbone need not be contracted and processed recursively.
Instead, the algorithm contracts only the maximal non-singleton sets that are not visited by the backbone.
The resulting strongly laminar instance together with the backbone forms a \emph{vertebrate pair}.

\begin{definition}[Vertebrate pair]
A \emph{vertebrate pair} consists of a strongly laminar instance $\cI = (G,T,\cL,x,y)$ of Asymmetric Multi-Depot TSP with $G=(V,E)$ and a connected Eulerian multi-graph $B$, the \emph{backbone}, with $V(B) \cap U \neq \emptyset$ for all $U \in \cL_{\geq 2} \coloneqq \{L \in \cL\colon |L| \geq 2 \}$.
We say that it is a \emph{simple} vertebrate pair if $|T| = 1$.

A \emph{$(\kappa,\eta)$-algorithm for vertebrate pairs} is a polynomial-time algorithm that computes an Eulerian multi-set $F \subseteq E$ of edges such that $(V, E(B) \cupp F)$ is connected and Eulerian, and
\begin{equation}\label{eq:kappa_eta_algo}
c^y(F) \ \leq\ \kappa \cdot \LP(\cI) \ +\ \eta \cdot \sum_{v \in V \setminus V(B)} 2y_v \cdot x(\delta^{-}(v)),
\end{equation}
where we set $y_v \coloneqq 0$ for $v \in V \setminus V(B)$ with $\{ v \} \not\in \cL$.
It is called a \emph{degree-preserving} $(\kappa,\eta)$-algorithm if the returned solution $F$ satisfies
\[ 
\floor{x(\delta^-(t))}
\ \leq\ |F \cap \delta^- (t)|
\ \leq\ \ceil{x(\delta^-(t))}
\]
for every $t \in T$.
\end{definition}

In \cite{traub25,vygen26}, the definition of a vertebrate pair requires $B$ to be a multi-subgraph of $G$.
However, the edges of $B$ are used only to ensure connectivity and do not contribute to the cost bound~\eqref{eq:kappa_eta_algo}.
Hence, we may safely remove this requirement.
Moreover, we include the additional factor $x(\delta^{-}(v))$ in \eqref{eq:kappa_eta_algo}, so that our definition also covers the unit-weight case, in which we set $y_v = 1/2$ even if $x(\delta^{-}(v)) > 1$.
For strongly laminar instances that satisfy the tightness conditions, this agrees with the usual definition, because $x(\delta^{-}(v)) = 1$ when $y_v > 0$.

\ifspringer\else\pagebreak\fi

The variant of Svensson's algorithm in \cite{vygen26} achieves the following guarantee for standard ATSP:

\begin{theorem}[\cite{vygen26}]\label{thm:kappa_eta_for_simple_vertebrate_pairs}
There is a $(2,4+2\sqrt2 + \epsilon)$-algorithm for simple vertebrate pairs and any constant $\epsilon > 0$.
\end{theorem}

For us, the backbone plays another important role, since it allows us to reduce the general multi-depot case to vertebrate pairs.
To see this, let $C_T$ be an artificial directed cycle with $V(C_T) = T$. 
Every maximal set $L \in \cL_{\geq 2}$ can be contracted to a single vertex and then, after uncontracting it, handled independently as an instance of classical ATSP.
Finally, if we obtain an Eulerian multi-set of edges $F$ such that $(V, E(C_T) \cupp F)$ is connected and Eulerian, then $F$ alone is a feasible multi-depot solution.

It therefore remains to give a degree-preserving $(\kappa, \eta)$-algorithm that handles such vertebrate pairs (which need not be simple).
To this end, we revisit \emph{subtour covers}.

\subsection{Subtour Covers}

Svensson's algorithm for vertebrate pairs maintains an Eulerian multi-set of edges $H$ and iteratively adds edges until $(V, E(B) \cupp H)$ is connected.
In each iteration, it adds a carefully selected set of edges from a \emph{subtour cover} and reduces the number of connected components.

Following the reduction described above, it suffices for us to consider vertebrate pairs $(\cI, B)$ with $\cI=(G,T,\cL,x,y)$, where $V(B)=T$ and $\cL$ consists only of singleton sets.
We therefore use the following adapted definition of a subtour cover.

\begin{definition}[Subtour cover]
\label{def:subtour_cover}
Let $(\cI,B)$, where $\cI = (G,T,\cL,x,y)$ and $G = (V,E)$, be a vertebrate pair with $V(B) = T$ and $\cL_{\geq 2} = \emptyset$, and let $H$ be an Eulerian multi-subset of $E[V \setminus V(B)]$.
A \emph{subtour cover} for $(\cI,B,H)$ is an Eulerian multi-subset $F$ of $E$ such that $F \cap \delta(W) \neq \emptyset$ for the vertex set $W$ of every strongly connected component of the graph $(V \setminus V(B), H)$.
\end{definition}

\begin{lemma}
\label{lem:subtour_cover}
Let $(\cI,B,H)$ be as in \Cref{def:subtour_cover}.
There is a polynomial-time algorithm that computes a subtour cover $F$ that satisfies
\begin{itemize}
\item
$c^y(F) \leq \LP(\cI) + \sum_{v \in V \setminus V(B)} 2y_v$,

\item
$\floor{x(\delta^-(t))} \leq |F \cap \delta^- (t)| \leq \ceil{x(\delta^-(t))}$\ for every $t \in T$, and

\item
$|F \cap \delta^- (v) | \leq 2 \cdot x(\delta^- (v))$\ for all $v \in V \setminus T$.
\end{itemize}
\end{lemma}

The proof of \Cref{lem:subtour_cover} follows directly from the proof of the graph subtour cover theorem in \cite{traub25} (Theorem~6.6), based on \cite{svensson15}.
The three properties are not explicitly stated there, but follow directly from this proof.
In particular, the circulation rounding preserves the degree bounds of $\floor{x(\delta^-(t))}$ and $\ceil{x(\delta^-(t))}$ at each terminal.

We now use \Cref{lem:subtour_cover} as the subtour-cover subroutine in Svensson's algorithm for vertebrate pairs (see \cite{traub25} for more details of this algorithm).
Using the improvements of \cite{vygen26}, this yields:

\begin{theorem}
\label{thm:kappa_eta_for_vertebrate_pairs}
Let $(\cI,B)$ be a vertebrate pair with $\cI = (G,T,\cL,x,y)$ as in \Cref{def:subtour_cover} and \Cref{lem:subtour_cover} (in particular, $\cL_{\geq 2} = \emptyset$).
There is a degree-preserving $(1,4+2\sqrt2 + \epsilon)$-algorithm for such vertebrate pairs for any constant $\epsilon > 0$.

Moreover, this holds even without requiring that the strongly laminar instance $\cI$ satisfies the conditions $x(\delta(U)) = 2$ and $U \subseteq V(G) \setminus T$ for all $U \in \cL$.
\end{theorem}

It remains to argue why the degree bounds of the subtour cover carry over to the final solution.
Before the final iteration, Svensson's algorithm adds only edges from $E[V \setminus V(B)]$, and hence does not affect the degrees of the terminals as we chose $V(B) = T$.
In the final iteration, the added edges incident to the backbone are exactly those contained in the subtour cover.
The algorithm may add further cycles that are not contained in the subtour cover.
However, these cycles can be chosen to lie within $G[V\setminus V(B)]$ and therefore do not affect terminal degrees (cf. step~(6c) of Algorithm~7.19 in~\cite{traub25} and the proof of Lemma~10 in~\cite{vygen26}).
Moreover, using the clean-up rule of \cite{vygen26}, we never delete edges incident to $V(B) = T$.
We conclude that the terminal-degree bounds of \Cref{lem:subtour_cover} also hold for the final solution.

\subsection{Proof of \Cref{thm:weighted_multi_depot_atsp}}

Let $(G,T,\cL,x,y)$ be a strongly laminar instance, and let $v, w$ be vertices contained in a common member of $\cL$.
Let $U \in \cL$ be minimal with the property that $v,w \in U$.
Since $G[L]$ is strongly connected for every $L \in \cL$, we can fix a \emph{nice} $v$-$w$-path~$P_{v,w}$ within $G[U]$ that enters every set $L \in \cL$ at most once and leaves every set $L \in \cL$ at most once (Lemma~7.5 in~\cite{traub25}).
For $W \in \cL$ with $u,v \in W$, we define
\[ 
D_W(u,v) \ \coloneqq\ \sum_{L \in \cL\colon u \in L \subsetneq W} y_L \ +\ \sum_{L \in \cL\colon v \in L \subsetneq W} y_L \ +\ c^y(P_{u,v}), \quad D_W\ \coloneqq\ \max_{u,v \in W} D_W(u,v).
\]
Note that, for any vertices $p,q \in W$, the value $2y_W + D_W$ bounds the contribution of sets $L \in \cL$ with $L \subseteq W$ to the cost of entering~$W$ at $p$ from outside $W$, following $P_{p,q}$ inside $G[W]$, and leaving $W$ at $q$.

We will use the following lemma as a black-box result to prove \Cref{thm:weighted_multi_depot_atsp}.

\begin{lemma}[Lemma 7.11 in \cite{traub25}]
\label{lem:recursive_vertebrate_pairs}
Let $(G,T,\cL,x,y)$ be a strongly laminar instance and $W \in \cL$.
Then, given an $(\kappa,\eta)$-algorithm for simple vertebrate pairs, there is a polynomial-time algorithm that returns an Eulerian multi-set $F \subseteq E(G[W])$ such that $(W,F)$ is connected and has cost at most
\begin{equation}\label{eq:recursive_vertebrate_pairs}
c^y(F) \ \leq \ (3\kappa + \eta + 2) \cdot \sum_{L \in \cL\colon L \subsetneq W} 2y_L \ -\ (\kappa + \eta) \cdot D_{W}.
\end{equation}
\end{lemma}

Finally, we prove \Cref{thm:weighted_multi_depot_atsp}, which we restate here for convenience.
Note that \Cref{thm:weighted_multi_depot_atsp} directly implies \Cref{thm:multi_depot_main}.

\ThmWeightedMultiDepotATSP*
\begin{proof}
Let $\epsilon > 0$.
We begin with the unit-weight case.
Let $(G,T,c)$ be an instance of Asymmetric Multi-Depot TSP with $G=(V,E)$ and $c(e) = 1$ for all $e \in E$.
We consider the strongly laminar instance $\cI \coloneqq (G_x, T, \cL, x, y)$ where $G_x$ is the support graph of $x$ and 
\[
\cL \ \coloneqq\ \parset{\{v\}\colon v \in V}, \quad y_v \ \coloneqq\ y_{\parset{v}} \ \coloneqq\ 1/2 \quad\text{for all }v \in V.
\]
Here, we do not require that the strongly laminar instance satisfies the conditions $x(\delta(U)) = 2$ and $U \subseteq V \setminus T$ for all $U \in \cL$.
Observe that $c^y(e) = c(e)$ for all $e \in E$.
Let $B$ be a directed cycle on $T$.
By \Cref{thm:kappa_eta_for_vertebrate_pairs} applied to the vertebrate pair $(\cI, B)$, we obtain a degree-preserving Eulerian multi-set $F \subseteq E(G_x)$ such that $(V, F \cupp E(B))$ is connected and Eulerian.
Removing the edges of $B$ from this tour yields a feasible degree-preserving Multi-Depot solution.
Moreover, its cost can be bounded by
\[ c(F) 
\ =\ c^y(F)
\ \overset{\eqref{eq:kappa_eta_algo}}{\leq}\  \LP(\cI) \ +\ (4 + 2\sqrt{2} + \epsilon) \cdot \sum_{v \in V \setminus T} 2y_v \cdot x(\delta^-(v))
\ \leq\ (5+ 2\sqrt{2} + \epsilon) \cdot c(x). \]

Next, we treat the general case.
To this end, set $\kappa \coloneqq 2$ and $\eta \coloneqq 4+ 2\sqrt{2} + \epsilon$.
By \Cref{lem:reduction_to_strongly_laminar}, it suffices to consider an arbitrary strongly laminar instance $\cI = (G,T,\cL,x,y)$ with $G=(V,E)$.
Let 
\[ \cL_{\max} \coloneqq \parset{U \in \cL\colon U \text{ maximal in } \cL } \]
be the set of maximal elements of $\cL$.
Consider the strongly laminar instance $\cI^\prime \coloneqq (G^\prime,T, \cL^\prime,x^\prime,y^\prime)$, where $G^\prime =(V^\prime, E^\prime)$ and $x^\prime$ result from $G$ and $x$ by contracting each $U \in \cL_{\max}$ into a single vertex $v_U$ and setting
\[
\cL^\prime \ \coloneqq\ \{ \{ v_U \}\colon U \in \cL_{\max} \}, \qquad y^\prime_{v_U} \ \coloneqq\ y_U \ +\ \frac{D_U}{2}.
\]
Again, let $B$ be a cycle on $T$ and apply \Cref{thm:kappa_eta_for_vertebrate_pairs} to the vertebrate pair $(\cI^\prime,B)$.
This yields an Eulerian multi-set $F^\prime \subseteq E^\prime$ of edges.
We transform this into an Eulerian multi-set $F \subseteq E$ of edges by replacing each traversal through a contracted vertex $v_U$, where $U \in \cL_{\max}$, with a shortest path between the corresponding entry and exit vertices.
Since the cost of this path can be bounded by $D_U$, we have
\begin{equation}\label{eq:tour_bound_global}
\begin{aligned}
c^y(F) 
\ \leq\ c^{y^\prime}(F^\prime) 
\ \overset{\eqref{eq:kappa_eta_algo}}&{\leq}\ \LP(\cI^\prime) \ +\ (4 +2\sqrt{2} + \epsilon) \cdot \sum_{U \in \cL_{\max}} 2y^\prime_{v_U} \\
\ &\leq\ (\kappa + \eta) \cdot \sum_{U \in \cL_{\max}} (2y_U + D_U).
\end{aligned}
\end{equation}

Next, apply \Cref{lem:recursive_vertebrate_pairs} (together with \Cref{thm:kappa_eta_for_simple_vertebrate_pairs}) to all $U \in \cL_{\max}$, and obtain multi-sets~$F_U \subseteq E$ of edges that each satisfy
\begin{equation}\label{eq:tour_bound_local}
c^y(F_U)
\ \overset{\eqref{eq:recursive_vertebrate_pairs}}{\leq}\ (3 \kappa + \eta + 2) \cdot \sum_{L \in \cL: L \subsetneq U} 2y_{L} \ -\ (\kappa + \eta) \cdot D_U.
\end{equation}
Observe that
$F \cupp\ \bigcupp_{U \in \cL_{\max}} F_U$ yields a feasible Multi-Depot TSP solution for our instance $\cI$.
Combining \eqref{eq:tour_bound_global} and \eqref{eq:tour_bound_local} shows that its cost can be bounded by
\begin{align*}
\sum_{U \in \cL_{\max}} \left( (\kappa + \eta) \cdot 2y_U \ +\ (3 \kappa + \eta + 2) \cdot \sum_{L \in \cL\colon L \subsetneq U} 2y_L \right)
\ &\leq\ (3\kappa + \eta + 2) \cdot \sum_{L \in \cL} 2y_L \\
\ &=\ (3\kappa + \eta + 2) \cdot c^y(x).
\end{align*}
Finally, this solution is degree-preserving, because each $U \in \cL_{\max}$ is contained in $V \setminus T$ and hence all terminal degrees are equal to those of $F^\prime$, which satisfies the required bounds.
\end{proof} 

\section{Asymmetric Single-Source/Single-Sink Multi-Path TSP}
\subsection{Equivalence to Other Variants}
\label{sec:equivalence_to_other_variants}

A natural variant to study is the \emph{Asymmetric Multiple-Source/Multiple-Sink Multi-Path TSP}.
Here, the instance contains terminal sets $S=\parset{s_1,\dots,s_k},T=\parset{t_1,\dots,t_k} \subseteq V$ such that $S \cap T = \emptyset$ and $|S|=|T| = k \geq 1$.
The task is then to find a permutation $\pi\colon [k] \to [k]$ and $s_i$-$t_{\pi(i)}$-walks $P_i$ for all $i \in [k]$ such that $V = \bigcup_{i = 1}^k V(P_i)$ and $\sum_{i=1}^k c(P_i)$ is minimized, or to decide that no such collection of walks exists.

A simple reduction, which was already observed by Friggstad~\cite{friggstad13}, shows that this variant is equivalent to the Single-Source/Single-Sink variant.

\begin{proposition}\label{prop:single_vs_multiple_sources}
Given an $\alpha$-approximation algorithm for Asymmetric Single-Source/\allowbreak\mbox{Single}-Sink Multi-Path TSP, we obtain an $\alpha$-approximation algorithm for Asymmetric Multiple-Source/Multiple-Sink Multi-Path TSP, and vice versa.
\end{proposition}
\begin{proof}

Let $(G,c,S,T)$ be an instance of Asymmetric Multiple-Source/Multiple-Sink Multi-Path TSP (with arbitrary weights).
Let $S = \{s_1, \dots, s_k\}$ and $T = \{t_1, \dots, t_k\}$.
Let $G^\prime$ result from $G$ by adding $2(k+1)$ new vertices $s^\ast, s^\prime_1, \dots, s^\prime_k$ and $t^\ast,t_1^\prime, \dots, t^\prime_k$, and zero-cost edges
\[
(s^\ast,s^\prime_i), (s^\prime_i, s_i), \quad\text{and} \quad (t_i, t^\prime_i), (t^\prime_i, t^\ast) \text{ for all $i \in [k]$}.
\]
Finally, observe that there is a canonical cost-preserving mapping between Multiple-Source/Multiple-Sink solutions for $(G,c,S,T)$ and Single-Source/Single-Sink solutions for~$(G^\prime,c,s^\ast,t^\ast,k)$.

For the reverse direction, given a Single-Source/Single-Sink instance $(G,c,s,t,k)$, we may replace the vertices~$s$ and $t$ by $k$ distinct co-located copies each and obtain an equivalent Multiple-Source/Multiple-Sink instance.
\end{proof}

Using a similar proof as for \Cref{prop:single_vs_multiple_sources}, we are able to prove \Cref{lem:initial_dp_solution}, which we restate below for convenience.

Recall that in the setting of \Cref{lem:initial_dp_solution}, we are given an instance $\cI = (G,c, \{(s_i,t_i) \}_{i=1}^k)$ of Asymmetric Multi-Path TSP with $G=(V,E)$ and $n = |V|$, and we denote $S \coloneqq \parset{s_1, \dots, s_k}$ and $T \coloneqq \parset{t_1, \dots, t_k}$.
Moreover, we assume without loss of generality that $c$ is metric and $G$ is the complete graph (by considering the metric closure).

\LemInitialDpSolution*
\begin{proof}
Construct $G^\prime$ from $G$ by adding two new vertices $s^\ast$ and $t^\ast$ and zero-cost edges $(s^\ast,v)$ and $(v,t^\ast)$ for all $v \in V(G)$.

Consider the instance $\cI^\prime \coloneqq (G^\prime,c,s^\ast,t^\ast,k)$ of Asymmetric SSSS Multi-Path TSP.
Since any feasible solution to $\cI$ induces a solution to $\cI^\prime$ of the same cost, we have $\OPT_{\cI^\prime} \leq \OPT_{\cI}$.
Then, by applying \Cref{thm:single_source_single_sink_main} with $\epsilon = 1$, we obtain $s^\ast$-$t^\ast$-walks $Q_1, \dots, Q_{\ell}$ for some $\ell \in \{k, \dots, 2k\}$ of total cost at most $51 \cdot \OPT_{\cI^\prime}$ (the approximation factor $51$ is given in the proof of \Cref{thm:single_source_single_sink_main}).
Finally, shortcutting repeated visits, as well as all visits to $s^\ast$, $t^\ast$, and vertices in $S \cup T$, we obtain the desired collection of paths.
\end{proof}

\subsection{Unit-weight Asymmetric SSSS Multi-Path TSP}
\label{sec:unweighted_ssss_multi-path_atsp}

In this subsection, we prove the second part of \Cref{thm:single_source_single_sink_main}.
That is, we give a $(5+2\sqrt{2}+\epsilon)$-approximation algorithm for the unit-weight Asymmetric SSSS Multi-Path TSP.

\ifspringer
\begin{proof}[second part of \Cref{thm:single_source_single_sink_main}]
\else
\begin{proof}[Proof of the second part of \Cref{thm:single_source_single_sink_main}]
\fi
Let $\epsilon > 0$ and $\mdatspconstant \coloneqq 5+ 2\sqrt{2} + \epsilon/2$. 
Let $(G,c,s,t,k)$ with $G=(V,E)$ be an instance of unit-weight Asymmetric SSSS Multi-Path TSP, i.e., $c(e) = 1$ for all $e \in E$.
We compute an optimal LP solution $x^\ast$ to the LP relaxation~\eqref{lp:multi_path_primal}.
Since $s \neq t$, we have $c(x^\ast) \geq k$.

Next, we construct an auxiliary graph $G^\prime$ (see \Cref{fig:graph_sequence}).
\begin{figure}[t]
\centering
\begin{tikzpicture}[
xscale=1.2,
yscale=1.1,
vertex/.style={circle,draw=black, fill,inner sep=1.5pt, outer sep=2pt},
edge/.style={-latex,very thick}
]
\definecolor{col1}{named}{black}
\definecolor{col2}{named}{black}
\definecolor{col3}{named}{black}
\definecolor{col4}{named}{blue} 

 \draw (0,0) ellipse [x radius=1.0cm, y radius=0.4cm];
 \node[] at (0, -0.7) {\scriptsize $G_1$};
  
 \draw (1.8,0) ellipse [x radius=1.0cm, y radius=0.4cm];
 \node[] at (1.8, -0.7) {\scriptsize $G_2$};
  
 \draw (3.6,0) ellipse [x radius=1.0cm, y radius=0.4cm];
 \node[] at (3.6, -0.7) {\scriptsize $G_3$};
   
 \draw (7.2,0) ellipse [x radius=1.0cm, y radius=0.4cm];
 \node[] at (7.2, -0.7) {\scriptsize $G_L$};

  \node[vertex,label=above:{\scriptsize$v^\ast$}] (v) at (3.6, 1) {};

\node[vertex,rectangle,label={[label distance=8pt]below:{\scriptsize $s_1$}},minimum size=1.2mm] (s1) at (-0.9,0) {};
\node[vertex,rectangle,label={[label distance=8pt]below:{\scriptsize $t_1=s_2$}},minimum size=1.2mm] (s2) at (0.9,0) {};
\node[vertex,rectangle,label={[label distance=8pt]below:{\scriptsize $t_2=s_3$}},  minimum size=1.2mm] (s3) at (2.7, 0) {};
\node[vertex,rectangle,label={[label distance=8pt]below:{\scriptsize $t_3=s_4$}},  minimum size=1.2mm] (s4) at (4.5, 0) {};
\node[vertex,rectangle,label={[label distance=8pt]below:{\scriptsize $t_{L-1}=s_L$}},  minimum size=1.2mm] (s5) at (6.3, 0) {};
\node[vertex,rectangle,label={[label distance=8pt]below:{\scriptsize $t_L$}},  minimum size=1.2mm] (s6) at (8.1, 0) {};
			
\draw[edge, red!80!black] (s6.north) to[out=145,in=0] (v.east);
\draw[edge, red!80!black] (v.west) to[out=180,in=35] (s1.north);  
\draw[edge, bend right=10, blue!80!black] (s1) to (s2);
\draw[edge, bend right=30, blue!80!black] (s1) to (s2);
\draw[edge, bend right=-10, blue!80!black] (s1) to (s2);
\draw[edge, bend right=-30, blue!80!black] (s1) to (s2);

\draw[edge, bend right=10, blue!80!black] (s2) to (s3);
\draw[edge, bend right=30, blue!80!black] (s2) to (s3);
\draw[edge, bend right=-10, blue!80!black] (s2) to (s3);
\draw[edge, bend right=-30, blue!80!black] (s2) to (s3);

\draw[edge, bend right=10, blue!80!black] (s3) to (s4);
\draw[edge, bend right=30, blue!80!black] (s3) to (s4);
\draw[edge, bend right=-10, blue!80!black] (s3) to (s4);
\draw[edge, bend right=-30, blue!80!black] (s3) to (s4);

\draw[edge, bend right=10, blue!80!black] (s5) to (s6);
\draw[edge, bend right=30, blue!80!black] (s5) to (s6);
\draw[edge, bend right=-10, blue!80!black] (s5) to (s6);
\draw[edge, bend right=-30, blue!80!black] (s5) to (s6);

\node[] at (5.4, 0) {$\dots$};

\end{tikzpicture} \vspace*{1.5em}
\caption{
The auxiliary graph $G^\prime$ we construct in the proof of the second part of \Cref{thm:single_source_single_sink_main}.
The artificial depot vertex $v^\ast$ is incident to two new edges (red), each traversed $k$ times in the Multi-Depot solution $F^\prime$.
Within each copy $G_i$, the restriction $F_i$ of that solution yields a feasible SSSS Multi-Path solution (blue).
}
\label{fig:graph_sequence}
\end{figure}
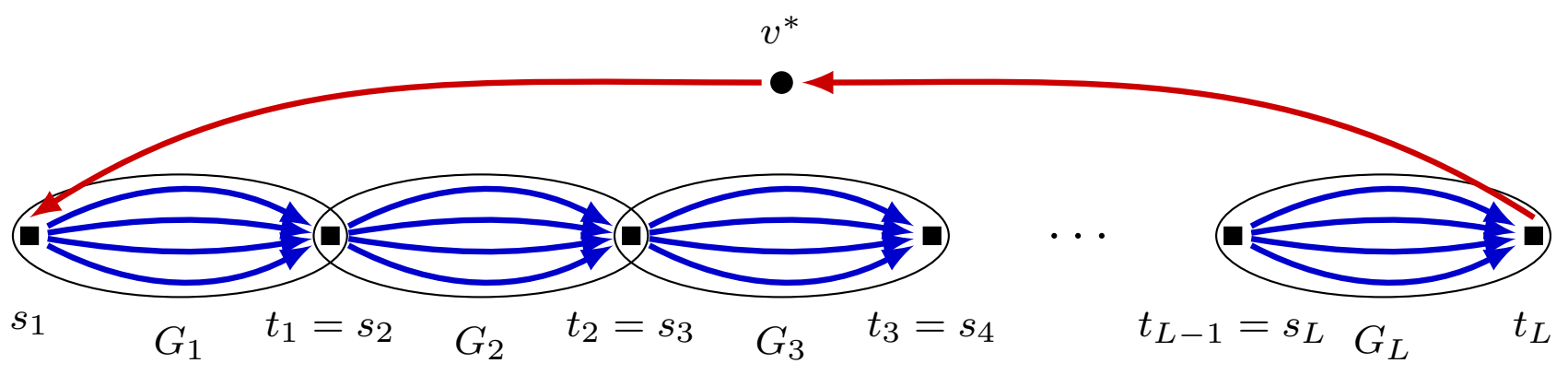
To this end, take $L \coloneqq \ceil{4\mdatspconstant/\epsilon}$ copies $G_1, \dots, G_L$ of $G$.
For $v \in V$, we denote the copy of $v$ in $G_i$ by $v_i$.
Then, for each $i \in [L-1]$, contract $t_i$ and $s_{i+1}$ into a single vertex.
Finally, add a new vertex $v^\ast$ and edges $(v^\ast, s_1)$ and $(t_L, v^\ast)$.
The resulting graph is $G^\prime$.
Moreover, we extend $x^\ast$ to $x^\prime$ by setting $x^\prime_{(v_i,w_i)} \coloneqq x^\ast_{(v,w)}$ for all $(v,w) \in E(G)$ and all $i \in [L]$,
and $x^\prime_{(v^\ast, s_1)} \coloneqq x^\prime_{(t_L, v^\ast)} \coloneqq k$.

Consider the instance $(G^\prime,T^\prime,c)$ of unit-weight Asymmetric Multi-Depot TSP, where $T^\prime \coloneqq \{ v^\ast \}$ (and $c \equiv 1$).
Observe that $x^\prime$ is a feasible solution to the Multi-Depot LP \eqref{lp:multi_depot_primal}.
Hence, we may apply \Cref{thm:weighted_multi_depot_atsp} to $x^\prime$ and obtain a degree-preserving Multi-Depot solution $F^\prime$ with $c(F^\prime) \leq \mdatspconstant \cdot c(x^\prime)$.

Let $F_i \coloneqq F^\prime \cap E(G_i)$ for all $i \in [L]$.
Note that, since $F^\prime$ is degree-preserving, we have $|\delta^-_{F^\prime}(v^\ast)|=|\delta^+_{F^\prime}(v^\ast)|=k$.
This implies $|\delta_{F_i}^+(s_i)| - |\delta_{F_i}^-(s_i)| = |\delta_{F_i}^-(t_i)| - |\delta_{F_i}^+(t_i)| = k$ for all $i \in [L]$.
In particular, each $F_i$ induces a feasible SSSS Multi-Path solution for~$(G,c,s,t,k)$.
We return the cheapest of these solutions, which has cost at most
\begin{align*}
\min_{i \in [L]} c(F_i) 
\ \leq\ \frac{c(F^\prime)}{L}
\ \leq\ \frac{\mdatspconstant \cdot c(x^\prime)}{L}
\ &=\ \frac{\mdatspconstant \cdot( L \cdot c(x^\ast) + 2k)}{L}	\\
\ &\leq\ \mdatspconstant \cdot c(x^\ast) + \frac{2\mdatspconstant  \cdot c(x^\ast)}{L} \\
\ &\leq\ (5+2\sqrt{2} + \epsilon) \cdot c(x^\ast).
\end{align*}
\end{proof}

\subsection{Proof of \Cref{lem:worst_case_gap}}
\label{sec:proof_worst_case_gap}

In this subsection, we prove \Cref{lem:worst_case_gap}, which we restate here for convenience.
\LemWorstCaseGap*

Our proof follows that of \cite{köhne20} almost verbatim.
Hence, we recall only the main arguments here and refer to their proof for further details.

Let $(G,c,s,t,k)$ with $G=(V,E)$ be an instance of Asymmetric SSSS Multi-Path TSP.
We compute in polynomial time an optimal dual LP solution $(a,y)$ to \eqref{lp:multi_path_dual} with the following properties:
\begin{enumerate}
\item
$a_s - a_t$ is minimum among all optimal dual solutions to \eqref{lp:multi_path_dual},

\item
$\cL \coloneqq \{U \subseteq V \colon y_U > 0 \}$ is laminar, and

\item
$G[U]$ is strongly connected for all $U \in \cL$.
\end{enumerate}
We refer to \cite{köhne20,traub25} for details on why such an optimal dual solution exists (and how to compute one in polynomial time).

Given this solution $(a,y)$, let $U_1, \dots, U_\ell \in \cL$ be pairwise disjoint, with $\ell \leq k$.
We claim that $t$ is reachable from $s$ in $G[V \setminus \overline{U}]$, where $\overline{U} \coloneqq \bigcup_{i=1}^\ell U_i$.
Suppose, for the sake of contradiction, that this is not the case.
Then define $R \subseteq V \setminus \{t\}$ to be the set of vertices reachable from $s$ in $G[V \setminus \overline{U}]$.
Obtain a dual solution $(\bar{a}, \bar{y})$ to \eqref{lp:multi_path_dual} by modifying $(a,y)$ as follows:
\begin{itemize}
\item
decrease $y_{U_i}$ by $\epsilon \coloneqq \min_{i \in [\ell]} y_{U_i}$ for all $i \in [\ell]$,

\item
decrease $a_v$ by $\epsilon$ for all $v \in R$, and

\item
increase $a_v$ by $\epsilon$ for all $v \in V \setminus (R \cup \overline{U})$.
\end{itemize}
Note that $t \in  V \setminus (R \cup \overline{U})$ by our assumption.
This yields a feasible dual solution to \eqref{lp:multi_path_dual}, since the value of $a_w - a_v +  \sum_{U\colon e \in \delta(U)} y_U $ only increases for edges from $R$ to $V \setminus (R \cup \overline{U})$, which do not exist by the definition of $R$.
Moreover,
\[
k(\bar{a}_t - \bar{a}_s) \ +\  \sum_{U \in \cL} 2\bar{y}_U 
\ =\  k(a_t - a_s) \ +\ 2k \epsilon \ +\ \sum_{U \in \cL} 2y_U \ -\ 2\ell\epsilon 
\ \geq\ \LP,
\]
implying that $(\bar{a}, \bar{y})$ is optimal.
In particular, $\bar{a}_s - \bar{a}_t < a_s - a_t$, contradicting the first property.

This claim and a variant of Menger's theorem yield $s$-$t$-paths $P_1, \dots, P_{k+1}$ such that no maximal set $U \in \cL$ is visited by two different paths.
Moreover, using the second and third properties of~$(a,y)$, these paths can be chosen so that
\begin{equation}
\label{eq:path_crossing_bound}
\sum_{i=1}^{k+1} \Big(|\delta^- (U) \cap E(P_i)| \ +\ |\delta^+ (U) \cap E(P_i)| \Big) \ \leq\ 2
\end{equation}
for all $U \in \cL$.
Recall our definition $c^y(e) \coloneqq \sum_{U\colon e \in \delta(U)} y_U$ for an edge $e$.
By complementary slackness and because we consider only edges in the support graph of an optimal primal solution to \eqref{lp:multi_path_primal}, we have
$c(e) = c^y(e) + a_w - a_v$ for all $e=(v,w) \in E$.
Altogether, we obtain that
\begin{align*}
0
\ \leq\ \sum_{i=1}^{k+1} c(P_i)
\ &=\ (k+1) \cdot (a_t - a_s) \ +\ \sum_{i=1}^{k+1} c^y(P_i) \\
\ \overset{\eqref{eq:path_crossing_bound}}&{\leq}\ (k+1) \cdot (a_t - a_s) \ +\ \sum_{U \in \cL} 2y_U.
\end{align*}
Multiplying both sides of this inequality by $k$ and adding $\sum_{U \in \cL} 2y_U$ to both sides yields
\begin{align*}
\sum_{U \in \cL} 2y_U \ \leq\ k(k+1) \cdot (a_t - a_s) \ +\ (k+1) \cdot \sum_{U \in \cL} 2y_U 
\ =\ (k+1) \cdot \LP,
\end{align*}
as claimed in \Cref{lem:worst_case_gap}.
We remark that this bound is tight; see \Cref{fig:worst_case_gap_tight} for an example.
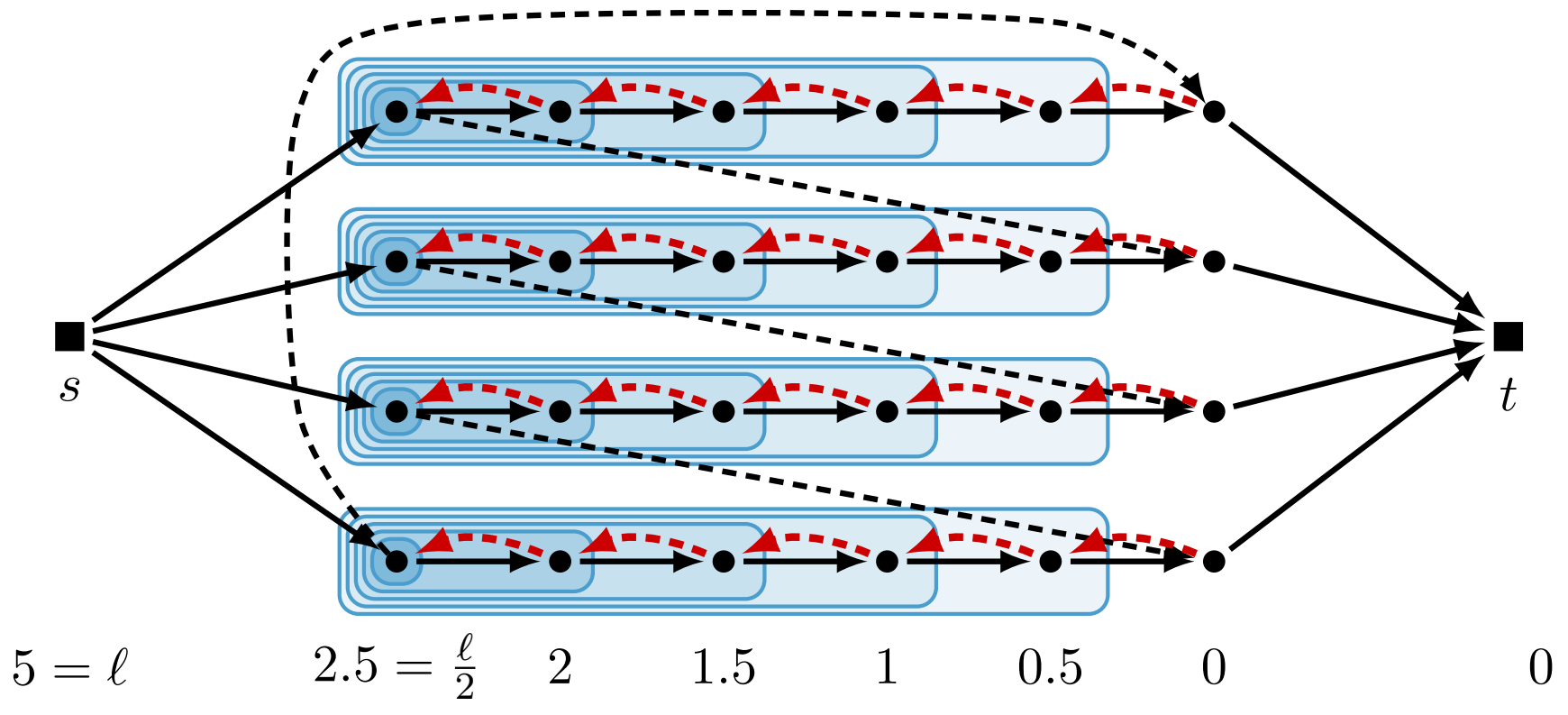
\begin{figure}[t]
\centering
\begin{tikzpicture}[
xscale=1.2,
yscale=1.1,
vertex/.style={circle,draw=black, fill,inner sep=1.5pt, outer sep=2pt},
edge/.style={-latex,very thick}
]
\definecolor{col1}{named}{black}
\definecolor{col2}{named}{black}
\definecolor{col3}{named}{black}
\definecolor{col4}{named}{blue}

\definecolor{regionblue}{HTML}{4A9DCC}

\tikzset{
box/.style={
thick,
rounded corners=4pt,
draw=regionblue,
fill=regionblue!70,
}
}

\draw[box,fill opacity=0.15] (0.65,0.65) rectangle (5.35,1.35);
\draw[box,fill opacity=0.15] (0.7,0.7) rectangle (4.3,1.3);
\draw[box,fill opacity=0.2] (0.75,0.75) rectangle (3.25,1.25);
\draw[box,fill opacity=0.4] (0.8,0.8) rectangle (2.2,1.2);
\draw[box,fill opacity=1] (0.85,0.85) rectangle (1.15,1.15);

\draw[box,fill opacity=0.15] (0.65,1.65) rectangle (5.35,2.35);
\draw[box,fill opacity=0.15] (0.7,1.7) rectangle (4.3,2.3);
\draw[box,fill opacity=0.2] (0.75,1.75) rectangle (3.25,2.25);
\draw[box,fill opacity=0.4] (0.8,1.8) rectangle (2.2,2.2);
\draw[box,fill opacity=1] (0.85,1.85) rectangle (1.15,2.15);

\draw[box,fill opacity=0.15] (0.65,2.65) rectangle (5.35,3.35);
\draw[box,fill opacity=0.15] (0.7,2.7) rectangle (4.3,3.3);
\draw[box,fill opacity=0.2] (0.75,2.75) rectangle (3.25,3.25);
\draw[box,fill opacity=0.4] (0.8,2.8) rectangle (2.2,3.2);
\draw[box,fill opacity=1] (0.85,2.85) rectangle (1.15,3.15);

\draw[box,fill opacity=0.15] (0.65,3.65) rectangle (5.35,4.35);
\draw[box,fill opacity=0.15] (0.7,3.7) rectangle (4.3,4.3);
\draw[box,fill opacity=0.2] (0.75,3.75) rectangle (3.25,4.25);
\draw[box,fill opacity=0.4] (0.8,3.8) rectangle (2.2,4.2);
\draw[box,fill opacity=1] (0.85,3.85) rectangle (1.15,4.15);

\node[vertex] (x11) at (1, 1) {};
\node[vertex] (x21) at (2, 1) {};
\node[vertex] (x31) at (3, 1) {};
\node[vertex] (x41) at (4, 1) {};
\node[vertex] (x51) at (5, 1) {};
\node[vertex] (x61) at (6, 1) {};

\draw[edge] (x11) to (x21);
\draw[edge] (x21) to (x31);
\draw[edge] (x31) to (x41);
\draw[edge] (x41) to (x51);
\draw[edge] (x51) to (x61);

\node[vertex] (x12) at (1, 2) {};
\node[vertex] (x22) at (2, 2) {};
\node[vertex] (x32) at (3, 2) {};
\node[vertex] (x42) at (4, 2) {};
\node[vertex] (x52) at (5, 2) {};
\node[vertex] (x62) at (6, 2) {};

\draw[edge] (x12) to (x22);
\draw[edge] (x22) to (x32);
\draw[edge] (x32) to (x42);
\draw[edge] (x42) to (x52);
\draw[edge] (x52) to (x62);

\node[vertex] (x13) at (1, 3) {};
\node[vertex] (x23) at (2, 3) {};
\node[vertex] (x33) at (3, 3) {};
\node[vertex] (x43) at (4, 3) {};
\node[vertex] (x53) at (5, 3) {};
\node[vertex] (x63) at (6, 3) {};

\draw[edge] (x13) to (x23);
\draw[edge] (x23) to (x33);
\draw[edge] (x33) to (x43);
\draw[edge] (x43) to (x53);
\draw[edge] (x53) to (x63);

\node[vertex] (x14) at (1, 4) {};
\node[vertex] (x24) at (2, 4) {};
\node[vertex] (x34) at (3, 4) {};
\node[vertex] (x44) at (4, 4) {};
\node[vertex] (x54) at (5, 4) {};
\node[vertex] (x64) at (6, 4) {};

\draw[edge] (x14) to (x24);
\draw[edge] (x24) to (x34);
\draw[edge] (x34) to (x44);
\draw[edge] (x44) to (x54);
\draw[edge] (x54) to (x64);

\node[vertex,rectangle,label=below:{$s$},  minimum size=2mm] (s) at (-1, 2.5) {};

\node[vertex,rectangle,label=below:{$t$},  minimum size=2mm] (t) at (7.8, 2.5) {};

\draw[edge] (s) to (x11);
\draw[edge] (s) to (x12);
\draw[edge] (s) to (x13);
\draw[edge] (s) to (x14);

\draw[edge] (x61) to (t);
\draw[edge] (x62) to (t);
\draw[edge] (x63) to (t);
\draw[edge] (x64) to (t);

\draw[edge, densely dashed, bend left=0] (x14) to (x63);
\draw[edge, densely dashed, bend left=0] (x13) to (x62);
\draw[edge, densely dashed, bend left=0] (x12) to (x61);
\draw[edge, densely dashed, bend left=0] plot[smooth] coordinates{(0.95,1.05)(0.4,2) (0.4,4)(1.2,4.6) (5,4.6) (5.95,4.05)};

\node[] at (-1, 0.3) {$5 = \ell$};
\node[] at (1, 0.3) {$2.5 = \frac{\ell}{2}$};
\node[] at (2, 0.3) {$2$};
\node[] at (3, 0.3) {$1.5$};
\node[] at (4, 0.3) {$1$};
\node[] at (5, 0.3) {$0.5$};
\node[] at (6, 0.3) {$0$};
\node[] at (8, 0.3) {$0$};

\draw[edge,red!80!black,densely dashed,bend right=25, ultra thick] (x63) to (x53);
\draw[edge,red!80!black,densely dashed,bend right=25, ultra thick] (x53) to (x43);
\draw[edge,red!80!black,densely dashed,bend right=25, ultra thick] (x43) to (x33);
\draw[edge,red!80!black,densely dashed,bend right=25, ultra thick] (x33) to (x23);
\draw[edge,red!80!black,densely dashed,bend right=25, ultra thick] (x23) to (x13);

\draw[edge,red!80!black,densely dashed,bend right=25, ultra thick] (x62) to (x52);
\draw[edge,red!80!black,densely dashed,bend right=25, ultra thick] (x52) to (x42);
\draw[edge,red!80!black,densely dashed,bend right=25, ultra thick] (x42) to (x32);
\draw[edge,red!80!black,densely dashed,bend right=25, ultra thick] (x32) to (x22);
\draw[edge,red!80!black,densely dashed,bend right=25, ultra thick] (x22) to (x12);

\draw[edge,red!80!black,densely dashed,bend right=25, ultra thick] (x61) to (x51);
\draw[edge,red!80!black,densely dashed,bend right=25, ultra thick] (x51) to (x41);
\draw[edge,red!80!black,densely dashed,bend right=25, ultra thick] (x41) to (x31);
\draw[edge,red!80!black,densely dashed,bend right=25, ultra thick] (x31) to (x21);
\draw[edge,red!80!black,densely dashed,bend right=25, ultra thick] (x21) to (x11);

\draw[edge,red!80!black,densely dashed,bend right=25, ultra thick] (x64) to (x54);
\draw[edge,red!80!black,densely dashed,bend right=25, ultra thick] (x54) to (x44);
\draw[edge,red!80!black,densely dashed,bend right=25, ultra thick] (x44) to (x34);
\draw[edge,red!80!black,densely dashed,bend right=25, ultra thick] (x34) to (x24);
\draw[edge,red!80!black,densely dashed,bend right=25, ultra thick] (x24) to (x14);

\end{tikzpicture}
 \vspace*{1.5em}
\caption{
An example for $k=3$ showing that the bound in \Cref{lem:worst_case_gap} is tight, adapted from an example due to Friggstad, Gupta, and Singh~\cite{friggstad16}.
It consists of $k+1$ layers, each containing $\ell+1=6$ vertices.
All black edges have cost $0$, and all red edges have cost $1$.
An optimal primal LP solution of total cost $\ell = 5$ assigns $x_e=\frac{k}{k+1}$ to every solid edge and $x_e=\frac{1}{k+1}$ to every dashed edge.
An optimal dual solution assigns $y_U=1/2$ to each subset $U$ shown in blue, with the values of $a_v$ indicated below the vertices (in particular, $a_s= \ell = 5$ and $a_t=0$).
Thus, $(k+1)\cdot\LP= (k+1)\ell =\sum_{\emptyset\neq U\subseteq V\setminus\parset{s,t}}2y_U$.
}
\label{fig:worst_case_gap_tight}
\end{figure}

\subsection{Asymmetric Regret-Bounded Vehicle Routing}
\label{subsec:regret_bounded_vrp}

Recall that in the \emph{Asymmetric Regret-Bounded Vehicle Routing Problem} we are given a complete digraph $G=(V,E)$ with a root vertex $r \in V$, edge costs $c\colon E \to \bR_{\geq 0}$ satisfying the triangle inequality, and a regret bound~$R \in \bR_{\geq 0}$.
The task is to find a minimum number of $r$-rooted paths $P_1, \dots, P_k$ such that every vertex is covered and each path $P_i$ has regret at most $R$, that is,
\[
\regret(P_i) \ \coloneqq\ c(P_i) \ -\ c(r,v_i) 
\ \leq\ R,
\]
where $v_i$ is the endpoint of $P_i$.

In this subsection, we prove \Cref{thm:regret_bounded_vrp}, which we restate here for convenience.
\ThmRegretBoundedVRP*

We prove \Cref{thm:regret_bounded_vrp} via the following black-box reduction to Asymmetric SSSS Multi-Path TSP:

\begin{theorem}[Friggstad and Swamy~\cite{friggstad14}]
\label{thm:regret_bounded_bicriteria_reduction}
Given an $(\alpha,\beta)$-bicriteria approximation algorithm for Asymmetric SSSS Multi-Path TSP, we obtain an $(\alpha + \beta)$-approximation algorithm for the Asymmetric Regret-Bounded Vehicle Routing Problem.
\end{theorem}

We remark that Friggstad and Swamy~\cite{friggstad14} (Theorem 6.6) state the approximation guarantee of this reduction as $O(\alpha + \beta)$. However, the constant of one can be directly derived from their proof.

Before proving \Cref{thm:regret_bounded_vrp}, we start by setting a baseline showing that any $\alpha$-approximation for ATSP yields a $(4\alpha + \epsilon)$-approximation for Asymmetric Regret-Bounded Vehicle Routing.

\ifspringer\else\pagebreak\fi

\paragraph{A simple black-box reduction to ATSP.}
Let~$\cI = (G,c,s,t,k)$ be an instance of Asymmetric SSSS Multi-Path TSP.
Add an edge $\overleftarrow{e} = (t,s)$ of cost 
\[
C \ \coloneqq\ \varphi \cdot \frac{\OPT_k}{k} 
\]
to $G$, where $\varphi > 0$ and $\OPT_k$ denotes the cost of an optimal solution to $\cI$. 
By standard scaling and rounding techniques, at the cost of a factor of $(1+ \epsilon)$ in the approximation guarantee, we may assume that our algorithm correctly guesses the value $\OPT_k$ for any constant $\epsilon > 0$.
We omit the details of this here.

This yields a corresponding instance of ATSP, for which a tour of cost at most $\OPT_k + kC = (\varphi + 1) \cdot \OPT_k$ exists.
Hence, an $\alpha$-approximation algorithm for ATSP returns a tour of cost at most
\[
\alpha (\varphi + 1) \cdot \OPT_k
\]
in this instance.
Moreover, this tour may use the artificially added edge $\overleftarrow{e}$ at most
\[
\frac{\alpha (\varphi + 1) \cdot \OPT_k}{C} \ = \ \frac{\varphi + 1}{\varphi} \cdot \alpha \cdot k
\]
times.\footnote{
It is also easy to ensure that this edge $\overleftarrow{e}$ is used at least $k$ times by subdividing it with a vertex and creating $k$ copies of the subdivided edge.
However, to obtain an approximation algorithm for Asymmetric Regret-Bounded Vehicle Routing it is not even necessary that $\overleftarrow{e}$ is used at least $k$ times.
}
Hence, we obtain an $(\alpha (\varphi + 1) + \epsilon, \frac{\varphi + 1}{\varphi} \cdot \alpha)$-bicriteria approximation for $\cI$.
By \Cref{thm:regret_bounded_bicriteria_reduction} and setting $\varphi \coloneqq 1$, this yields a $(4\alpha + \epsilon)$-approximation algorithm for Asymmetric Regret-Bounded Vehicle Routing.
Using the approximation guarantees of \cite{vygen26} and \cite{hajiaghayi26} for ATSP, we obtain approximation ratios less than $60$ and $50$, respectively.

\paragraph{Improving the approximation guarantee.}
In the proof of the first part of \Cref{thm:single_source_single_sink_main}, we obtain a $(\frac{51}{\epsilon}, 1+\epsilon)$-bicriteria approximation for Asymmetric SSSS Multi-Path TSP for any $0 < \epsilon \leq 1$.
In particular, this yields a $53$-approximation for Asymmetric Regret-Bounded Vehicle Routing.
However, this can be improved.

Let $(G,c,s,t,k)$ with $G=(V,E)$ be an instance of Asymmetric SSSS Multi-Path TSP and let $(a,y)$ be an optimal solution to the linear program \eqref{lp:multi_path_dual}, and denote its value by $\LP$.
Moreover, let $(a^\prime,y^\prime)$ be an optimal solution to \eqref{lp:multi_path_dual} for the instance $(G,c,s,t,k^\prime)$, where $k^\prime \coloneqq (1+\delta) k$ with $\delta \coloneqq 4$, and denote its value by $\LP^\prime$.
As in the proof of \Cref{lem:lp_path_scaling}, we have
\begin{equation}\label{eq:LP_dual_bound}
\LP 
\ \geq \ k A^\prime \ +\ 2Y^\prime,
\end{equation}
where $A^\prime \coloneqq a^\prime_t - a^\prime_s$ and $Y^\prime \coloneqq \sum_{\emptyset \neq U \subseteq V \setminus \parset{s,t}} y^\prime_U$.
Recall that $\LP^\prime = k^\prime A^\prime + 2 Y^\prime$.
Then,
\begin{align*}
kA^\prime \ +\ 2Y^\prime
\ &=\ \frac{1}{1 + \delta} \cdot k^\prime A^\prime \ +\ 2Y^\prime \\
\ &=\  \frac{1}{1 + \delta} \cdot (\LP^\prime - 2Y^\prime) \ +\ 2Y^\prime \\ &=\ \frac{\delta}{1 + \delta} \cdot 2Y^\prime \ +\ \frac{1}{1 + \delta} \cdot \LP^\prime.
\end{align*}
Combining this with \eqref{eq:LP_dual_bound} shows
\[
2Y^\prime \ \leq\ \frac{1 + \delta}{\delta} \cdot \LP \ -\ \frac{1}{\delta} \cdot \LP^\prime.
\]
Let $\mdatspconstant \coloneqq 12 + 2\sqrt{2} + \frac{1}{100}$ (\Cref{thm:weighted_multi_depot_atsp}).
Note that $\mdatspconstant - 1 \geq \delta$.
Then, by applying \Cref{lem:dual_relative_algo} to $(G,c,s,t,k^\prime)$ and $(a^\prime, y^\prime)$, we obtain a solution consisting of at most $k^\prime$ walks.
Their total cost is at most
\begin{align*}
k^\prime A^\prime \ +\ \mdatspconstant \cdot 2Y^\prime
\ &\leq\ \LP^\prime \ +\ (\mdatspconstant - 1) \cdot \left( \frac{1+\delta}{\delta} \cdot \LP \ -\ \frac{1}{\delta} \cdot \LP^\prime \right) \\
\ &\leq \ (\mdatspconstant - 1) \cdot \frac{1+ \delta}{\delta} \cdot \LP. 
\end{align*}
Recall that $\delta = 4$.
Using \Cref{thm:regret_bounded_bicriteria_reduction}, we therefore obtain an approximation guarantee of
\[
(\mdatspconstant - 1) \cdot \frac{1+ \delta}{\delta}  \ +\ (1 + \delta) 
\ \leq \ 22.3
\]
for Asymmetric Regret-Bounded Vehicle Routing. 
\end{document}